\documentclass[fleqn,usenatbib]{mnras}

\usepackage{newtxtext,newtxmath}
\usepackage[T1]{fontenc}

\usepackage[normalem]{ulem}

\usepackage{verbatim}

\DeclareRobustCommand{\VAN}[3]{#2}
\let\VANthebibliography\thebibliography
\def\thebibliography{\DeclareRobustCommand{\VAN}[3]{##3}\VANthebibliography}

\newcommand{\el}[2]{\ensuremath{^{#1}\mathrm{#2}}}

\usepackage{graphicx}	% Including figure files
\usepackage{amsmath}	% Advanced maths commands
\usepackage{hyperref}
\usepackage{lipsum}
\usepackage{csquotes}

\title[\text{Inhomogeneous GCE Models for LMC UFDs}]{Inhomogeneous Galactic Chemical Evolution: Modelling Ultra-Faint Dwarf Galaxies of the Large Magellanic Cloud}

\author[Alexander et al.]{
R. K. Alexander,$^{1,2,5}$\thanks{E-mail: r.alexander-2021@hull.ac.uk}
F. Vincenzo,$^{1,5}$
A. P. Ji, $^{3,5}$
H. Richstein, $^{4}$
C. J. Jordan, $^{1}$
and B. K. Gibson$^{1,5}$
\\ ~ \\
$^{1}$E.A. Milne Centre for Astrophysics, University of Hull, Hull, HU6~7RX, United Kingdom\\
$^{2}$Centre of Excellence for Data Science, AI, and Modelling (DAIM), University of Hull, Cottingham Road, Kingston-upon-Hull, HU6 7RX\\
$^{3}$Department of Astronomy \& Astrophysics, University of Chicago, 5640 S Ellis Avenue, Chicago, IL, 60637, USA\\
$^{4}$Department of Astronomy, University of Virginia, 530 McCormick Road, Charlottesville, VA, 22904, USA \\ 
$^{5}$Joint Institute for Nuclear Astrophysics, Center for the Evolution of the Elements (JINA-CEE)
}
\date{Accepted 2023 April 27. Received 2023 April 27; in original form 2023 March 06}

\pubyear{2022}

\begin{document}
\label{firstpage}
\pagerange{\pageref{firstpage}--\pageref{lastpage}}
\maketitle

\begin{abstract}
Ultra-faint dwarf galaxies are among the oldest and most metal-poor galaxies in the cosmos, observed to contain no gas and a high dark matter mass fraction. Understanding the chemical abundance dispersion in such extreme environments could shed light on the very first generations of stars. We present a novel inhomogeneous chemical evolution model, {\tt i-GEtool}, that we apply to two ultra-faint dwarf galaxies, Carina II and Reticulum II, both satellites of the Large Magellanic Cloud. Our model is based on the Monte Carlo sampling of the initial mass function as star formation proceeds in different gas cells of the galaxy volume. We account for the chemical enrichment of Supernova bubbles as they spread in the interstellar medium, causing dispersion in the elemental abundances. We recreate the abundance patterns of $\alpha$- and odd-$\textit{Z}$ elements, predicting two sequences in [C/Fe] and [N/Fe] at all metallicities. Our models underestimate [C/Fe] and [Ti/Fe] because of the large uncertainty in the adopted stellar nucleosynthesis yields. We discuss that the observed C and N abundances had likely been affected by internal mixing processes, which changed the initial surface abundances in the red giants. Our Supernova feedback scheme is responsible for driving galactic outflows, which quench the star formation activity at early times. We predict an average outflow mass-loading factor $\approx 10^{3}$, which extrapolates towards very low galaxy stellar masses the trend observed at high masses. Finally, by combining our model with the MIST isochrone database, we compare our synthetic colour-magnitude diagrams to observations.
\end{abstract}


%%%%%%%%%%%%%%%%%%%%%%%%%%%%%%%%%%%%%%%%%%%%%%%%%%

%%%%%%%%%%%%%%%%% BODY OF PAPER %%%%%%%%%%%%%%%%%%

\section{Introduction}

Ultra-Faint Dwarf galaxies (UFDs) are among the most dark matter dominated galaxies in the cosmos (e.g., see \citealt{battaglia2022_nature}), observed to contain a total number of stars of the order $10^{3} \lesssim \text{N}_{*} \lesssim 10^{5}$ and mass-to-light ratios as high as $\approx 10^{3}\,\big( M/L \big)_{\sun}$ \citep{hayashi2022}; as such they are difficult to observe, having very low surface brightness \citep{belokurov2007, Simon2007, koch2009, mcconnachie2012, Bechtol2015, simon2019, belokurov2022}. Currently, there are $\approx$ 21 spectroscopically confirmed UFDs scattered around the Milky Way (MW) halo, with a further 21 UFDs without detailed dynamical analysis yet \citep{simon2019,cerny2022}. The vast majority of the UFDs belong to the MW halo with approximately half a dozen of them being linked to the Large Magellanic Cloud (LMC) \citep{kallivayalil2018, erkal2020, patel2020, sacchi2021}. 

Dwarf galaxies have been historically classified into two main categories, those with gas (also known as dwarf irregulars) and those without (dwarf ellipticals), depending mainly on their distance to the parent galaxy and the physical conditions of the circumgalactic environment, with dwarf ellipticals being more clustered than dwarf irregulars (e.g., see \citealt{vader1991}). From the analysis of H$\scriptstyle\mathrm{I}$ survey observational data, \citet{Putman2021} were able to measure how the gas content of dwarf galaxy satellites changes, on average, as a function of their distance to the central galaxy, finding that the dwarf galaxies at distances $d < 2$ Mpc away from the MW and M31 contain no detectable H$\scriptstyle\mathrm{I}$ gas (see also \citealt{grcevich2009}). In agreement with this finding, the UFD galaxy satellites of the MW and LMC do not contain any detectable gas at the present time. 

Several scenarios have been postulated to explain how gas might have left UFD galaxies, which we briefly summarise in what follows (see also the review of \citealt{collins2022}, and references therein).
\textit{(i)} Supernova (SN) explosions can drive intense outflows of gas from the shallow potential well of dwarf galaxies, effectively quenching their star formation history and reducing the star formation efficiency \citep{dekel1986,Gibson1997, dekel2003, lanfranchi2003, lanfranchi2004,salvadori2009,vincenzo2014, romano2019,Gallart2021}. \textit{(ii)} Another scenario proposes that in the early Universe, when UFD galaxies formed, reionization removed most of the cold gas from their shallow dark matter halo, quenching their star formation activity (e.g., see \citealt{bullock2001,Katz2020}). 
\textit{(iii)} Another mechanism that could be responsible for the loss of gas of dwarf galaxies, quenching their star formation histories at early times, is ram pressure stripping \citep{gunn1972, ferguson1994, Mayer2006, grcevich2009, kirby2013, Putman2021}. \textit{(iv)} Finally, tidal stripping can also efficiently remove gas from the dwarf galaxy gravitational potential without the need for an external gas medium, as the mass loss would only be driven by tidal interactions (e.g., see \citealt{mayer2001, Mayer2006, fattahi2018} and references therein). All aforementioned mechanisms likely played a crucial role in the evolution of the gas content within UFD galaxies, with each process dominating at different evolutionary phases, depending also on the environment and the past orbital motion of UFD galaxies (e.g., \citealt{ricotti2005,salvadori2009}). 

Interestingly, from the analysis of the colour-magnitude diagram (CMD), \citet{sacchi2021} found that the UFD satellites of the LMC had, on average, their star formation quenched $\approx$ 600 Myr after the MW UFD satellites stopped forming stars.

In this work, we focus on modelling the chemical evolution of two spectroscopically confirmed UFDs -- Carina II (Car II) and Reticulum (Ret II), both satellites of the LMC \citep{Bechtol2015, patel2020, Battaglia2022}. Car II - along with Carina III - is generally believed to be very understood from the analysis in \citet{Li2018} and \citet{Ji2020}, whereas Ret II is chosen from its unique $r$-process enrichment early in its formation. We select these two satellite candidates among other UFDs for our analysis as their formation history is well recorded, as well as their chemical abundance measurements.

\begin{enumerate}

\item Car II is a satellite of the LMC and was discovered by the Magellanic SatelLites Survey (MagLites) while exploring some of the unknown regions around the Magellanic system \citep{Torrealba2018}. Car II is estimated to have a stellar mass $M_{\star} \approx 10^{4}\,\text{M}_{\sun}$ \citep{Sales2017}. Interestingly, \citet{Battaglia2022} analysed the bulk kinematics and spatial distributions of the Local Group dwarf galaxies along with the LMC and determined that Car II is likely a satellite of the LMC \citep{patel2020, erkal2020}. Further investigations into this hypothesis by \citet{Pace2022} concluded with similar findings that Car II was previously associated with the LMC and is on its first infall into the MW halo. Analysis of the elemental abundances within Car II confirmed that this galaxy followed a similar chemical evolution as the MW UFDs, and there is no sign of anti-correlation between light element abundances, confirming that Car II stars had not been part of any disrupted globular cluster \citep{Ji2020}.

\item Ret II was discovered by \citet{Bechtol2015} from the analysis of the Dark Energy Survey (DES) data. It has an estimated stellar mass $M_{\star}\approx 10^{3}\,\text{M}_{\sun}$ \citep{Sales2017, sacchi2021}. By fitting the observed CMD with the PARSEC library of stellar isochrones \citep{Bressan2012}, \citet{Bechtol2015} predicted that Ret II stellar populations have an average age $\tau \approx 13.5$ Gyr and metallicity $Z \approx 10^{-4}$, making this UFD galaxy one of the oldest and most metal-poor structures in the known Universe. Ret II was soon classified as a MW satellite \citep{Simon2015, Walker2015}; however, from the analysis of its orbital properties, \citet{erkal2020} and \citet{patel2020} determined that Ret II is gravitationally bound to the LMC and is experiencing its first infall into the MW along with the LMC. Concerning chemical abundance studies, \citet{Ji2020} used the \textit{Magellan Inamori Kyocera Echelle} (MIKE) spectrograph on the Magellan Clay telescope \citep{bernstein2003} to collect spectra of nine red giant-branch stars (RGB) within Ret II, being able to measure the abundances of several chemical elements, which we include in our work. 

\end{enumerate}

Here we developed a novel inhomogeneous chemical evolution model, {\tt i-GEtool}, which is built upon a previous model by \citet{Argast2000,Argast2002}, that we apply to investigate the chemical evolution of Ret II and Car II and recover their stellar properties. Chemical evolution models have been used for decades to investigate the chemical enrichment history of the stars and interstellar medium (ISM) of galaxies (e.g., see the books of \citealt{matteucci2012,pagel2009}). 
An ideal inhomogeneous chemical evolution model should, in principle, account for the stochastic formation of SN progenitors and be able to characterise how the nucleosynthetic products of stars disperse in the ISM as a result of SN explosions and stellar winds, mimicking also SN bubbles. 

One of the first inhomogeneous chemical evolution models in the literature was developed by \citet{Argast2000,Argast2002}, who studied how the ISM elemental abundances and their spread evolve as a function of time in the Galactic halo, by including an element of stochasticity in the model in the core-collapse SN enrichment. Later, \citet{G.Cescutti} developed an alternative inhomogeneous, stochastic chemical evolution model that was used to investigate the origin of the abundance spread of s- and r-process elements in MW halo stars. In \citet{G.Cescutti}, the stellar yields of neutron-capture elements from core-collapse SNe were tuned to reproduce how the observed elemental abundance spread changes as a function of metallicity, providing a powerful method to empirically constrain the stellar nucleosynthetic yields of neutron-capture elements from massive stars and other progenitors (see also \citealt{cescutti2013, cescutti2014, cescutti2015} for a similar approach, in which additional nucleosynthesis sources such as merging neutron stars are investigated in detail). Similarly, \citet{Wehmeyer2017} adapted the inhomogeneous chemical evolution model of \citet{Argast2000,Argast2002}, developing the code {\tt ICE}, to reproduce the chemical abundances of r-process elements as measured in Galactic halo stars, by focusing on different possible nucleosynthesis channels, including merging neutron stars.

Our work is organized as follows. Section \ref{section:2} provides a detailed description of our inhomogeneous chemical evolution model, {\tt i-GEtool}. The observed sample of elemental abundances as measured in the red giants of Car II and Ret II is presented in Section \ref{sec: 4}. Section \ref{section: 3} summarizes the main assumptions of our chemical evolution models for Car II and Ret II and shows the model predictions for their SFH, metallicity distribution function (MDF), age-metallicity relation, and mass-loading factor. Section \ref{sec:ce} goes through the chemical abundance patterns as predicted by our chemical evolution models for Car II and Ret II, exploring two separate yield tables, followed by Section \ref{sec:cmd}, in which we present our predictions for the CMD of Car II and Ret II. Lastly, the conclusions of our work are summarised in Section \ref{conclusion}. Throughout the paper we assume a flat $\Lambda$ cold dark matter (CDM) universe with the following cosmological parameters: {$\text{H}_0$} = 70 {$\text{kms} ^ {-1}$} {$\text{Mpc} ^ {-1}$}; {${\Omega} _ {\Lambda}$} = 0.73; {${\Omega} _ {M}$} = 0.27.

\section{Inhomogeneous Chemical Evolution Model}
\label{section:2}

In this Section, we present our inhomogeneous chemical evolution model, {\tt i-GEtool}, which has been developed to track how the elemental abundances in the stars and ISM of galaxies evolve as a function of time. We apply this model to predict the formation of two UFD satellite galaxies of the LMC, Car II and Ret II. Our model is originally based on the previous inhomogeneous chemical evolution model of \citet{Argast2000,Argast2002}, using similar parametrizations of various astrophysical processes.

\subsection{Initial Setup}

The simulation takes place within a three-dimensional box, which is divided into $40^3$ equally sized cells, each with sides $20\,\text{pc}$. Within this volume, we place primordial gas of H and He with abundances $X=0.74$ and $Y=0.26$ in mass, respectively, to recreate the environment of the early Universe from the epoch of reionization. At the start of the simulation, the gas in the simulated volume is assumed to have the following initial mass density profile as a function of radius:

\begin{equation}
    \rho_{\text{gas}}\big(r, t = 0\big) = a \times b^{-\frac{r}{200\,\text{pc}}},
    \label{density_profile}
\end{equation}

\noindent where $\rho_{\text{gas}}$ corresponds to the gas mass density at a distance $r$ (in pc) from the centre of the box, and $a=55\,\text{M}_{\sun} \, \text{pc}^{-3}$ and $b=85$ represent free parameters tuned to have $\approx 3 \times 10^{5} \,\text{M}_{\sun}$ of primordial gas mass within the galaxy at the start of the simulation. 

\subsection{Inflow}
\label{section:2.1}

In the simulation, the galaxy evolves by accreting primordial gas (mostly H and He) from the environment. Gas inflow is uniformly distributed throughout the volume and drives the formation of stars and hence the chemical enrichment of the ISM. We follow the accretion equation shown in \citet{kobayashi2020} where they describe how the total infall rate of gas is assumed to evolve as a function of time:

\begin{equation}
    \Big( \frac{d\,\rho_{\text{gas}}(t)}{dt} \Big)_{\text{infall}} = \mu \cdot t \cdot \exp({\frac{-t}{\lambda}}),
    \label{infall}
\end{equation}

\noindent where $t$ is the simulation time, and $\mu$ and $\lambda$ are free parameters chosen to vary the intensity and duration of the gas accretion rate respectively. Table \ref{table:prop} contains the values for $\lambda$ for our reference models of Car II and Ret II. The total gas mass which is accreted in the model for Car II and Ret II are $3.5 \times 10^{7} \, \text{M}_{\sun}$ and $1.1 \times 10^{7} \, \text{M}_{\sun}$ respectively, which constrains the normalization constant, $\mu$, of the infall rate in equation \ref{infall}.

\subsection{Star Formation}
\label{section:2.2}

The star formation rate (SFR) is assumed to follow a Schmidt-Kennicutt relation as follows. 

\begin{equation}
\label{SFE}
    \text{SFR}(t) = \epsilon \times \big\langle \rho(t) \big\rangle \, V,
\end{equation}

\noindent where $\epsilon$ is the so-called star formation efficiency (SFE), a free parameter of the model with units corresponding to the inverse of a time (see Table \ref{table:prop}), $\big\langle \rho(t) \big\rangle$ is the average gas mass density within the box at the time $t$, and $V$ is the total simulated galaxy volume. 

At a given time step $\Delta t$, stars are formed in the simulation by randomly sampling the \citet{Kroupa2001} initial mass function (IMF) until the total mass of stars that form within $\Delta t$ equals the quantity $\text{SFR}(t) \, \Delta t$. The position of the star-forming cells is selected at random in the UFD galaxy volume according to the following probability distribution, $P_{\text{SF}}$:
\begin{equation}
    P_{\text{SF}} \propto \rho_{\text{gas}}^{1.8},
    \label{equation: dw}
\end{equation}
\noindent where $\rho_{\text{gas}}$ corresponds to the total gas mass within each cell.
We assume that stars can only form within cells that have a total gas mass $M_{\text{thr}}>100\,\text{M}_{\sun}$, with such threshold corresponding to the maximum possible stellar mass that can be formed during each star formation event; in other words, the Monte Carlo sampling of the IMF is not performed (hence star formation does not take place) if the gas mass available within the cell is lower than the maximum stellar mass that can be formed. 

When star formation takes place in a cell, the newly-born star acquires mass at the expense of the adjacent eight cells. The chemical composition of the newly-born star is also inherited from the eight nearest cells, with the contribution from each cell being weighted by its gas mass normalized to the total mass of the eight cells combined together.

Once a star with mass $m$ is formed, its stellar lifetime, $\tau_{\star}$, is given by the following formula:

\begin{equation}
\begin{split}
    \log\Big( \frac{\tau_{\star}}{\text{Gyr}} \Big) = \alpha + \beta \cdot \log\Big( \frac{m}{\text{M}_{\sun}} \Big)
    + \gamma \cdot \log^2\Big( \frac{ m }{\text{M}_{\sun}}\Big),
    \label{equation: time}
\end{split}
\end{equation}

\noindent which was derived by \citet{Larson1974}, where $\alpha$, $\beta$ and $\gamma$ are 1.0, -3.42 and 0.88 respectively from \citet{Kobayashi2004} and \citet{David1990} (see, also, \citealt{johnson2021}). We deviate from the previous stellar lifetime equations derived by \citet{Argast2000} from the analysis of the Geneva stellar evolution models at different metallicities \citep{Schaller1992,Schaerer1993a,Schaerer1993b,Charbonnel1993}. This is to ensure that our very low mass stars ($m \leq 0.8$) are alive at 13.8 Gyr at the end of the simulation. 

\subsection{Stellar Nucleosynthetic Yields}
\label{section:2.3}

At the end of their lifetimes, stars enrich the ISM with their nucleosynthetic products. Low- and intermediate-mass stars ($0.8 \lesssim m \lesssim 8\,\text{M}_{\sun}$) contribute to the chemical evolution of galaxies through stellar winds when they reach the asymptotic giant branch (AGB) phase, producing chemical elements such as \el{12}{C} and \el{14}{N} \citep{Karakas2010}. Massive stars ($m \gtrsim 8\,\text{M}_{\sun}$) explode as core-collapse SNe (CCSNe), producing the majority of $\alpha$-elements such as O, Ne, Mg, Si, S, and Ca in the cosmos \citep{Kobayashi2006}. All stars with mass $m \lesssim 0.9\,\text{M}_{\sun}$ have typical lifetimes longer than the age of the Universe, hence they do not contribute with any nucleosynthetic product to the chemical enrichment of the ISM. Finally, Type Ia Supernovae (SNe Ia) are the main producers of iron-peak elements in the cosmos \citep{Kobayashi2006,Iwamoto1999}. 

In our model, we adopt the AGB stellar yields of \citet{Karakas2010}, which are available for masses in the range $1 \leq m \leq  6\,\text{M}_{\sun}$ and the following metallicities: $Z = 0.0001$, $0.004$, $0.008$, and $0.02$.  For massive stars, we adopt the stellar yields of \citet{Kobayashi2006}, including also their hypernova models. The stellar yields of \citet{Kobayashi2006} are available for masses in the range $13 \leq m \leq 40\,\text{M}_{\sun}$ and the following metallicities: $Z = 0.0$, $0.001$, $0.004$, and $0.02$. We assume that hypernovae contribute by $50$ per cent with respect to the global chemical enrichment of massive stars, with such fraction being the same for all metallicities. This fraction was chosen by \citet{Kobayashi2006} to reproduce [$\alpha$/Fe] and [Zn/Fe] observations in the Solar neighbourhood. Since the stellar yields of \citet{Kobayashi2006} are only available for stars with masses ranging $13\,\text{M}_{\sun} < m < 40\,\text{M}_{\sun}$, we use linear extrapolation to acquire values for initial masses higher than $40\,\text{M}_{\sun}$. Finally, we assume the SN Ia yields of \citet[W7 model]{Iwamoto1999}. 

To explore the intrinsic systematic uncertainty as due to different stellar models and nucleosynthesis calculations, we also develop chemical evolution models with an alternative set of stellar yields from the NuGrid collaboration \citep{Ritter2018}. The NuGrid stellar yields are available for masses in the range $1 \leq m \leq 25\,\text{M}_{\sun}$ and the following metallicities: $Z=0.0001$, $0.001$, $0.006$, $0.01$, and $0.02$.

\subsection{Delay-Time Distribution}
\label{section:3.5}

The explosion time of Type Ia SNe is acquired by making use of a mathematical formalism based on the so-called delay-time distribution (DTD) function \citep{Ruizlapuente1998}. In essence, the function $\text{DTD}(\tau)$ describes the probability of having a delay-time, $\tau$, that elapses between the formation of the  SN Ia progenitor binary system and the final SN Ia explosion. 

Recent studies of the DTD find that a power law of the form $N_{\text{Ia}}\,\tau^{-s}$ best reproduces the observed average Type Ia SN rate in galaxies as a function of the delay-time $\tau$, where $N_{\text{Ia}}$ is a normalisation constant and the slope $s$ varies from work to work in the literature, depending on the SN sample \citep{Maoz2010,Maoz2017,Chen2021}. For example, \citet{Maoz2017} measured a slope $s = -1.1 \pm 0.1$ and a normalization constant $N_{\text{Ia}} = (1.3 \pm 0.1) \times 10^{-3} \, \text{M}^{-1}_{\sun}$ ($N_{\text{Ia}}$ corresponds to the number of SNe Ia that are expected to explode for every $10^{3}\,\text{M}_{\sun}$ of stellar mass formed). A similar slope was measured by \citet{Castrillo2021}, who found $s = -1.1 \pm 0.3$, determining also a minimum delay-time $\tau_{\text{min,Ia}} = 50 ^{+100}_{-35}\,\text{Myr}$ from the formation of the Type Ia SN progenitor system. Interestingly, \citet{Maoz2010} measured the DTD of Type Ia SNe in the LMC, by dissecting the LMC disc in several cells and comparing the number of SN remnants in each cell to its past SFH. \citet{Maoz2010} found that the population of Type Ia SNe in the LMC could be explained by adopting a normalization $N_{\text{Ia}} = (2.7 \text{-} 11.0) \times 10^{-3} \,\text{M}_{\sun}^{-1}$ and an average delay-time $\tau_{\text{min,Ia}} = 330 \,\text{Myr}$ for the prompt Type Ia SNe.

Following the most recent findings as obtained by observational surveys of Type Ia SNe \citep{Maoz2010}, in our model we adopt the following formula for the DTD: 

\begin{equation}
\label{DTD_Ia}
    \text{DTD}_{\text{Ia}}(\tau) \, = \, N_{\text{Ia}}\,\tau^{-1.4} \;\;\;\text{for}\; \tau > \tau_{\text{min,Ia}},
\end{equation}

\noindent where the normalization constant, $N_{\text{Ia}}$, represents to the SN Ia probability (see Section \ref{section:2.4}), and $\tau_{\text{min,Ia}}$ represents the minimum delay-time. In our models, $N_{\text{Ia}}=8.0 \times 10^{-3} \, \text{M}_{\sun}^{-1}$ and $\tau_{\text{min,Ia}} = 152\,\text{Myr}$ for both Car II and Ret II. Specifically, \citet{brown2019} found that the specific SNe Ia rates are systematically higher in dwarf galaxies than larger ones (see also \citealt{Johnson2022} for a theoretical perspective). Recent works of \citet{Chen2021} predicts a different slope of -1.4, from which we adopt in our models.

\begin{table}
\begin{tabular}{|p{3.0cm}|p{2.0cm}|p{2.0cm}|}
    \hline
    Properties & Car II & Ret II \\
    \hline
    $\lambda$ $[\text{Myr}]$: & 550 & 100 \\
    $\epsilon$ [$\text{Gyr}^{-1}$]: & 0.006 & 0.004 \\
    \hline
    Number of Stars & $1.4 \times 10^{5}$ & $2.4 \times 10^{4}$ \\
    Stellar Mass [M$_{\sun}$] & $4.2 \times 10^{4}$ & $7.0 \times 10^{3}$ \\
    N$_{\text{CCSN}}$(t < t$_{\text{SF,max}}$) & 715 & 145 \\
    N$_{\text{SNIa}}$(t < t$_{\text{SF,max}}$) & 179 & 57 \\ 
    N$_{\text{CCSN}}$(t > t$_{\text{SF,max}}$) & 6 & 6 \\
    N$_{\text{SNIa}}$(t > t$_{\text{SF,max}}$) & 821 & 174 \\
    \hline
\end{tabular}

\caption{Properties of each model, highlighting some of the key differences, with the columns corresponding to Car II and Ret II. \textit{Row 1}: The peak accretion as a function of time. \textit{Row 2}: The stars formation efficiency used in Equation \ref{SFE}. \textit{Row 3}: The total number of stars in the model as predicted at the present day. \textit{Row 4}: The predicted stellar mass of the models at the present day. \textit{Row 5}: The number of core-collapse events which occurred during the star formation period. \textit{Row 6}: The number of Type 1a events which occurred during the star formation period. \textit{Row 7}: The number of core-collapse events which occurred after the star formation period. \textit{Row 8}: The number of Type 1a events which occurred after the star formation period.}
\label{table:prop}
\end{table}

\subsection{Supernova Explosions}
\label{section:2.4}

Every time a star with a mass in the range $0.9 \leq m \leq  8\,\text{M}_{\sun}$ is created by the Monte Carlo sampling of the IMF, a random number $r$ is uniformly drawn between $0$ and $1$. If $r<N_{\text{Ia}}$, namely if $r$ is lower than the SN Ia probability, then the star is assumed to be in a binary system that will give rise to a Type Ia SN explosion, with a delay-time that is drawn by randomly sampling the assumed DTD (see Section \ref{section:3.5}). Conversely, if $r\geq N_{\text{Ia}}$, we assume that the star will enrich the ISM during the AGB phase, by injecting its chemical products directly into the cell in which it was born.

All stars with mass $m>10\,\text{M}_{\sun}$ are assumed to end their lives as CCSNe. In our model, all supernova events (including SNe Ia) have the exact same explosion energy, $\text{E}_{\text{SN}} = 10^{51} \text{ergs}$, with the blast wave sweeping out $\text{M}_{\text{swept}} = 5 \times 10^{4.5} \; \text{M}_{\sun}$ of ISM mass -- being a higher value than $10^{4} \, \text{M}_{\sun}$ as assumed, for example, by \citet{Argast2002} so as to expel gas with higher efficiency in UFD galaxies. $\text{M}_{\text{swept}}$ can be redistributed in either a spherical or aspherical fashion to mimic the impact of asymmetric SN explosions; this creates a large hot diffuse medium, surrounded by a thin shell of dense gas which approximately corresponds to a \enquote*{bubble} of gas. For the purpose of redistributing the mass, all SNe that explode within a given cell are placed in the bottom-left corner of the cell. This simplifies our calculation when it comes to selecting the neighbouring cells within the SN explosion boundary.

To define the shape and size of the \enquote*{bubble} of gas resulting from a SN explosion, we draw a sphere which is centred on the SN location and initially comprises eight equally-sized octants, corresponding to the eight nearest eight cells to the SN explosion. Each octant steps out, cell by cell, counting the gas mass as it goes. When this count reaches $\text{M}_{\text{swept}}$ at a radius $\text{r}_{\text{swept}}$ centred on the SN location, the code transfers $90$ per cent of the gas from all cells at radii $\text{r} < \text{r}_{\text{swept}}$ into the cells at $\text{r} = \text{r}_{\text{swept}}$. This allows us to test the impact of asymmetric SN explosions as we generate inhomogeneous chemical enrichment in the simulated galaxy.

Our model does not incorporate gas mixing, as each cell remains separate and does not interact with others in its entirety. In a future study, we plan to allow the gas cells to acquire some gas and metals from the neighbour cells, to quantify the effects of mixing in the chemical evolution of UFDs and other galaxies. With an incorporated mixing coefficient added to the models, we predict the chemical dispersion within a single stellar population would decrease, leading to more stars having identical abundances throughout the simulation. Incorporating $r$-process elements such as Ba and Eu into our Ret II model would provide further clarity to the formation history and chemical evolution of the galaxy and build upon the works of \citet{ji2022}.

\begin{figure}
    \centering
    \includegraphics[width = 0.45\textwidth]{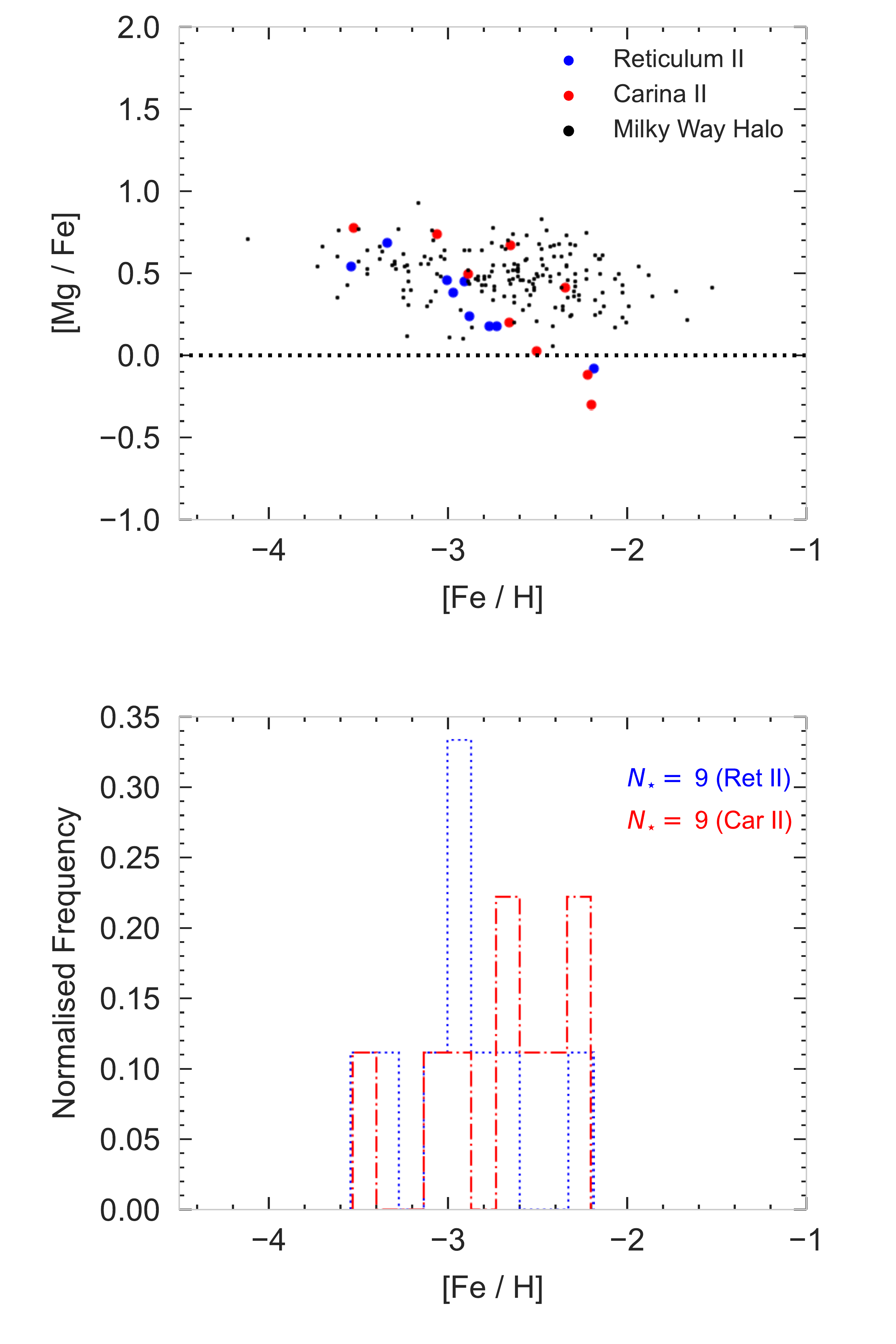}
    \caption{\textit{Top panel} -- The observed [Mg/Fe]-[Fe/H] chemical abundance pattern in Car II (red points from \citealt{Ji2020}) and Ret II (blue points from \citealt{Ji2016}) as compared to the observations in MW halo stars (black points from the LAMOST survey; \citealt{Li2022}). \textit{Bottom panel} -- The stellar MDF as observed in Car II (red dash-dotted line) and Ret II (blue dotted line) from the same sample of red giants as in the top panel. Both MDFs assume a bin width of $0.2\,\text{dex}$ and are normalised to unity.}
    \label{fig:OP}
\end{figure}

\subsection{Outflow}

In our model, galactic outflows of gas are assumed to be driven by SN explosions. Since the SN bubbles are three-dimensional, they can expand beyond the volume of the box; any SN bubble doing so is assumed to be part of the galactic outflow, and its swept-out mass $\text{M}_{\text{swept}}$ is lost from the system. This approach allows us to easily quench the star formation history in the model.

When enough gas leaves the galaxy and there is no cell left with gas mass $>100\,\text{M}_{\sun}$, the SFR is zero everywhere in the system, as there is not enough gas in the cells to sustain star formation by sampling the IMF.  After the end of the star formation activity, Type Ia SN explosions continue in the system, removing further enriched gas from the simulated volume. Our method of gas removal is unique to our model and serves as a key difference to \citet{Argast2000,Argast2002}.

\section{Observation Summary}
\label{sec: 4}

\begin{figure*}
    \centering
    \includegraphics[width = 1.00\textwidth]{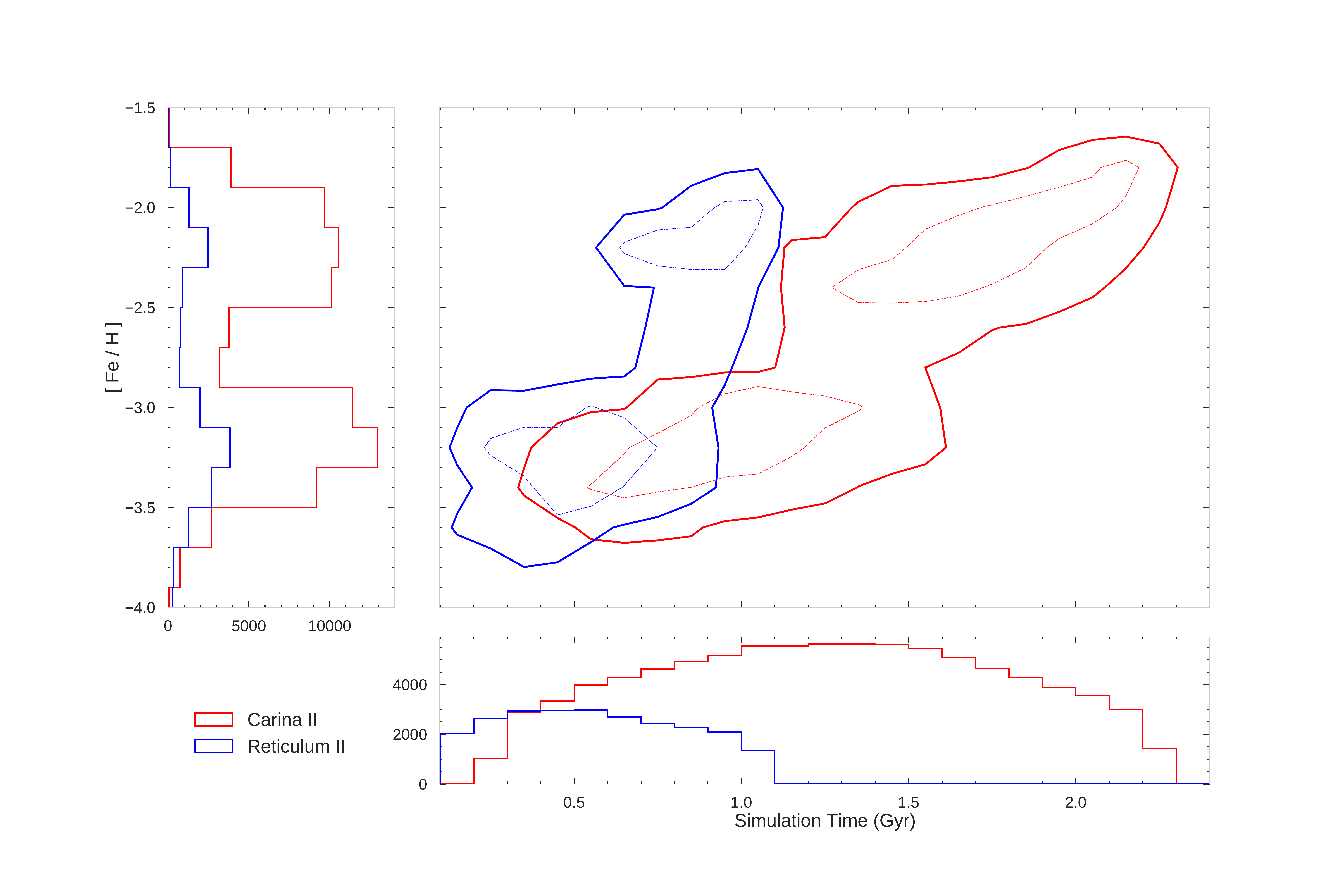}
    \caption{\textit{Middle panel} -- The predicted age-metallicity relation in Car II (red contours) and Ret II (blue contours). The contours correspond to densities of $15$ per cent (thick solid lines) and $50$ per cent (thin dashed lines) normalised to the maximum value of each distribution. \textit{Bottom panel} -- The predicted age distribution of the stars formed in Car II (red histogram) and Ret II (blue histogram), adopting a bin width of $0.1 \, \text{Gyr}$. \textit{Left panel} -- The predicted MDF in Car II (red histogram) and Ret II (blue histogram), adopting a bin width of $0.2\,\text{dex}$.}
    \label{fig:Car_PS}
\end{figure*}

The observational data of chemical abundances in Car II and Ret II is taken from the works of \citet{Ji2020} and \citet{Ji2016} respectively, who obtained spectra for nine stars in Car II with MagLiteS and nine stars in Ret II with the MIKE spectrograph on the Magellan Clay telescope. In the bottom panel of Figure \ref{fig:OP}, the metallicity distribution function (MDF) as observed in Car II (red histogram) is compared with the MDF in Ret II (blue histogram); both distributions are normalised to one. In the top panel of Figure \ref{fig:OP}, the [Mg/Fe] - [Fe/H] observations in Car II (red-filled circles) are compared with those in Ret II (blue-filled circles); in the same figure, we also show the chemical abundance measurements of [Mg/Fe] vs [Fe/H] in a sample of $385$ MW halo stars (black filled circle) as derived by \citet{Li2022} from the analysis of spectra collected with the Large sky Area Multi-Object fibre Spectroscopic Telescope (LAMOST). Additional constraints for our analysis are also provided by the observed CMDs (see Section \ref{sec:cmd}), which -- for Car II -- are based on observations from \citet{Li2018} that were obtained with the Magellan Baade Telescope, the Anglo-Australian Telescope, and the Very Large Telescope whereas for Ret II we use the observations that were performed with the Hubble Space Telescope (HST) as part of HST-GO 14734 (PI: Kallivayalil) and were presented by \citet{sacchi2021}. 

Though the observed sample is sparse and only consists of nine stars per galaxy, the red giants in Car II have systematically higher [Mg/Fe] ratios than those in Ret II at fixed [Fe/H] (see Fig. \ref{fig:OP}, top panel). Moreover, the red giants in Car II also have their [Fe/H] abundances concentrated towards higher values than in Ret II (Fig. \ref{fig:OP}, bottom panel). This seems to suggest that the star formation activity in Car II was more intense than in Ret II, producing a higher number of stars ($M_{\star,\text{CarII}}\approx 4 \times 10^{4}\,\text{M}_{\sun}$ whereas $M_{\star,\text{RetII}}\approx 7 \times 10^{3}\,\text{M}_{\sun}$;  see also \citealt{Sales2017}) and causing Type Ia SNe to explode when the ISM was not much polluted yet with iron from core-collapse SNe. This way, the chemical enrichment of Fe from Type Ia SNe produces a decrease of [Mg/Fe] in Ret II which takes place at lower [Fe/H] abundances than in Car II (see also \citealt{matteuccigreggio1986,matteuccibrocato1990,lanfranchi2004,vincenzo2014} for some further illustrations of this process).

\section{Modelling Carina II and Reticulum II}
\label{section: 3}

Here we present a detailed account of the star formation history (SFH) and chemical evolution as predicted by our inhomogeneous chemical evolution models for Car II and Ret II. The adopted parameters of the models are summarised in Table \ref{table:prop} and were obtained by developing a fine grid of models with the aim of reproducing the observed chemical abundance patterns along with the gas and stellar mass content as observed at the present time (see Section \ref{sec: 4}). Additional constraints for our analysis are provided by the observed CMDs, which will be discussed in Section \ref{sec:cmd}. 

\begin{figure}
    \centering
    \includegraphics[width = 0.45\textwidth]{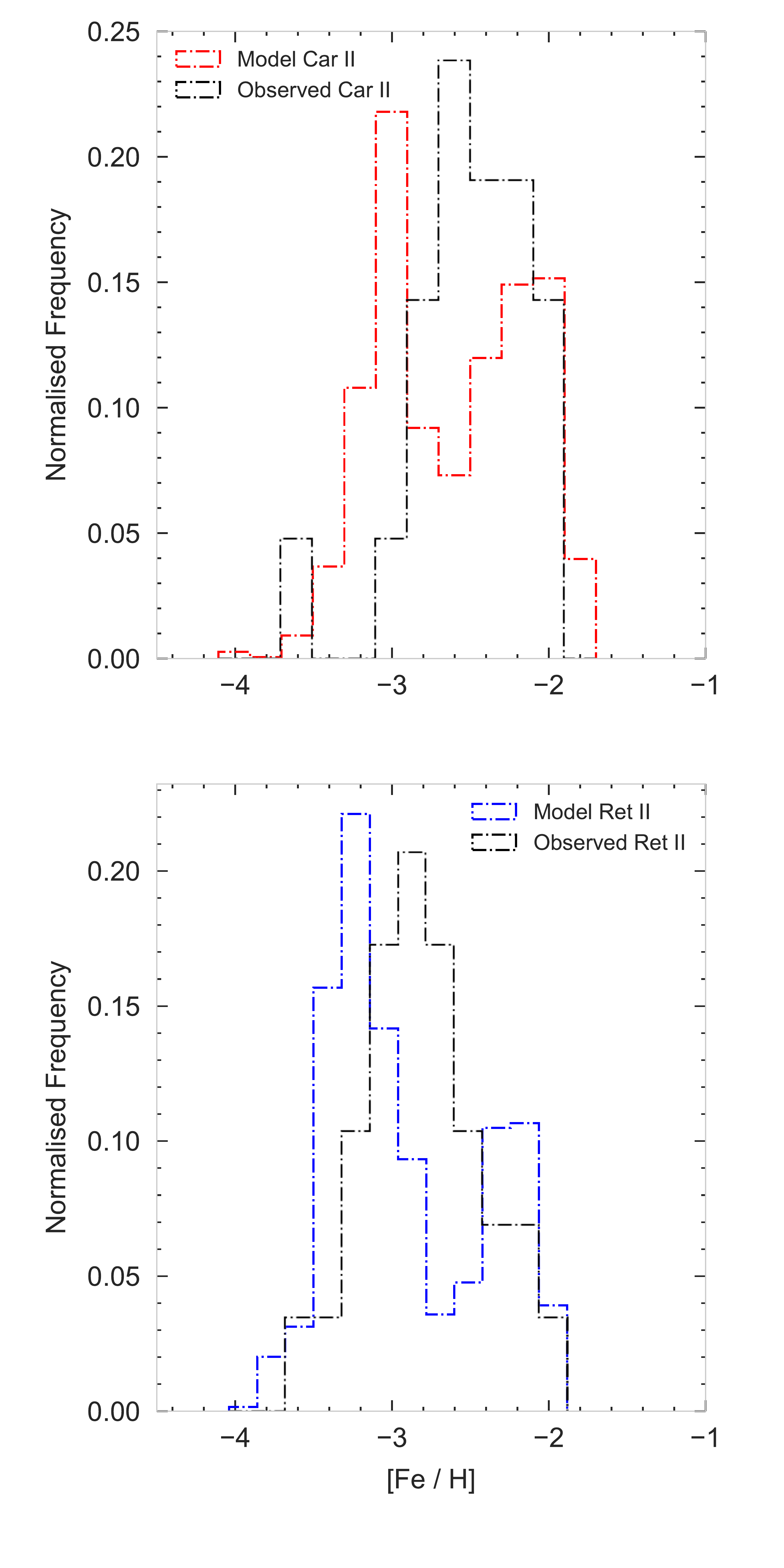}
    \caption{Comparison between the MDF of observed RGB stars (black dotted lines) and MDF of all model stars (red and blue dotted lines) in Car II (top panel) and Ret II (bottom panel), as labelled. In each panel, both distributions are normalised to unity, adopting a bin width of $0.2 \, \text{dex}$.}
    \label{fig:normal_cmd_comp}
\end{figure}

We remind the readers that the parameter $\mu$ in equation \ref{infall} for the infall rate is the normalisation constant which determines the total amount of gas which is accreted within the galaxy, while $\lambda$ is related to the accretion time-scale and regulates the time of peak accretion; varying these parameters drastically changes the SFH and chemical evolution as predicted by the model, as the gas content within each cell of the simulated volume determines the intensity of the star formation rate (see equation \ref{SFE}) and hence the rate of chemical enrichment of the ISM from dying stars and SNe.

Our Car II and Ret II models were derived from a detailed parameter study to reproduce the MDF and chemical abundances of observed stars. We varied the total gas accretion into the galaxy within parameter $\mu$, the accretion time-scale, $\lambda$, and the star formation efficiency, $\epsilon$. One of the main parameters we found to have a major impact on the MDF and chemical abundances is $\lambda$. In particular, higher $\lambda$ values produce longer star formation histories given a constant $\epsilon$ and $\mu$. Some other parameters which we adjusted and tuned are the normalisation constant, $N_{\text{Ia}}$ and the minimum delay time, $\tau_{\text{min,Ia}}$. We found that an increasing $N_{\text{Ia}}$ allowed more gas to be expelled from the galaxy, but will truncate formation quicker as a consequence. Similarly, an increasing $\tau_{\text{min,Ia}}$ would push back iron contributions in the chemical abundances of stars from SNe Ia and in turn extend the formation time period.

Reproducing the low stellar mass and metallicity content within Car II and Ret II requires very low star formation efficiencies (the quantity $\epsilon$, in Table \ref{table:prop}) of the order $5\times 10^{-3}\,\text{Gyr}^{-1}$, in agreement with the findings of previous studies \citep{salvadori2009,vincenzo2014}. Increasing the star formation efficiency, $\epsilon$, not only impacts the number of stars in the galaxy but also the number of supernova events, that are crucial in quenching the star formation history at early times. The chemical abundances for both Car II and Ret II are similar suggesting their gas reservoirs were diminished and star formation history quenched. Similarly, the minimum delay time for Type Ia SNe, $\tau_{\text{min,Ia}}$, and the normalization of the SN Ia rate, $N_{\text{Ia}}$, are both crucial to reproduce the observed elemental abundance patterns, as they regulate the time of onset of SNe Ia and the number of SNe Ia per unit mass of stellar mass formed, respectively, thus determining the chemical evolution of iron in the stars and ISM as a function of time. Car II is predicted to contain more stars and thus a higher stellar mass than Ret II, yet both galaxies show signs of a similar star formation history, concentrated at very early times. 

On the one hand, Ret II is characterised by a shorter and less intense SFH, which results from a short accretion time-scale ($\lambda_{\text{RetII}}=100\,\text{Myr}$) and a low star formation efficiency ($\epsilon_{\text{RetII}}=0.004\,\text{Gyr}^{-1}$), producing a large number of metal-poor stars. On the other hand, the best-fitting model for Car II is characterised by a more elongated SFH, which results from a longer accretion time-scale than in Ret II ($\lambda_{\text{CarII}}=550\,\text{Myr}$), allowing enough time for stars to evolve to the point where we get a similar MDF as the observed one. Different stellar nucleosynthetic yields in the model only impact the chemical evolution and abundances of the galaxy and not its SFH.

\citet{Ji2020} estimated the chemical enrichment from $\approx 10^{2}$ CCSNe is needed to explain the observed chemical abundances in Car II along with their spread. Even with the Monte Carlo sampling of the IMF, the creation of high-mass stars can be maintained and controlled in our chemical evolution model by varying the gas accretion history, the SFE and the maximum mass for CCSNe progenitors -- being $100\,\text{M}_{\sun}$ in our model. For Car II, we obtain a total number of $721$ CCSNe and $1000$ SNe Ia (see Table \ref{table:prop}). During the star formation period, we predict $715$ CCSNe and $179$ SNe Ia; after star formation ended, there are $821$ further SN Ia events in Car II which removed gas out of the galaxy. The number of SN events in Ret II is lower by a factor of 2 because of the less intense star formation activity within the system.

\begin{figure}
    \centering
    \includegraphics[width = 0.45\textwidth]{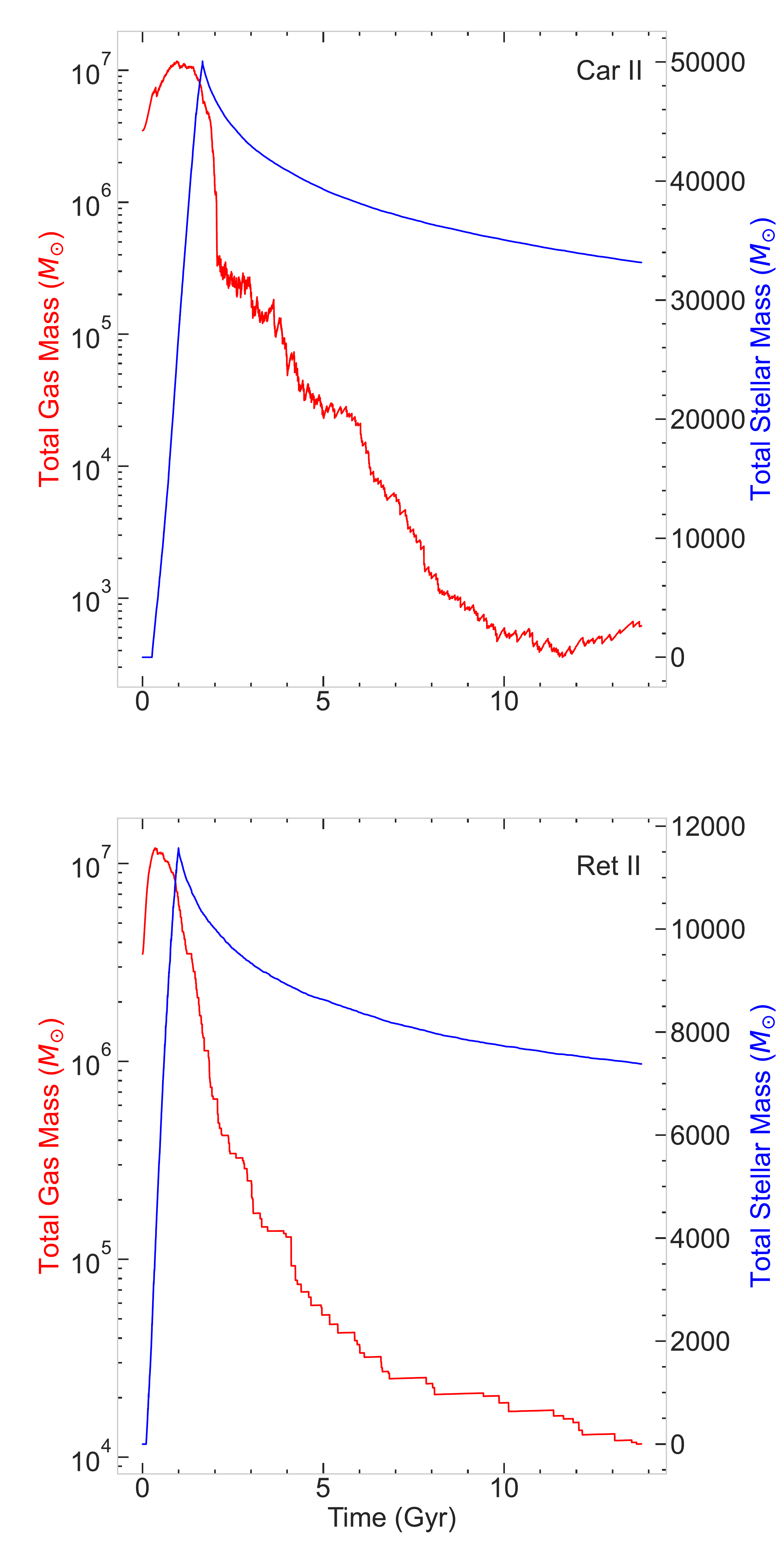}
    \caption{Predicted evolution of the total gas mass (red solid line) and total stellar mass (blue solid line) as a function of time as predicted in Ret II (top panel) and Car II (bottom panel). The model predictions are shown until the SFR is quenched by the SN-driven galactic outflows.}
    \label{fig:TGM}
\end{figure}

Figure \ref{fig:Car_PS} shows the stellar age-metallicity relation (AMR) as predicted by our reference chemical evolution models for Ret II (blue distribution) and Car II (red distribution); [Fe/H] is used as a proxy for the stellar metallicity. The contours in the figure correspond to densities of $15$ and $50$ per cent (thick solid lines and thin dashed lines, respectively) normalized to the maximum value of each distribution. Along with the AMR are the MDF on the left panel and the SFH on the bottom. As we are using the $\Lambda$CDM standard cosmological model, the assumed age of the Universe is 13.8 Gyr - which is the zero-point in time for the evolution of our simulated galaxy.

Both Car II and Ret II are satellites of the LMC, but their formation and evolution are predicted to be vastly different from one another, with Car II forming its stars during a more intense burst of star formation than Ret II. Stronger gas infall within our model for Car II also ensures that the gas content within the galaxy exceeds the threshold for star formation at an earlier time than in Ret II. This explains why the predicted SFHs as shown in the bottom panel of Fig. \ref{fig:Car_PS} have different evolution at early times. Both models predict a similar metallicity spread in their stellar populations at the present time, with the Car II model showing a steeper increase of [Fe/H] as a function of age when considering the oldest stellar populations, which stems from a more intense chemical enrichment activity at early times as due to massive stars and SNe Ia than in the model for Ret II.

In Figure \ref{fig:normal_cmd_comp}, the MDF of 21 stars in Car II as observed by \citet{Li2018} and \citet{Ji2020} (upper panel) and the MDF of 29 stars in Ret II from \citet{Simon2015} and \citet{ji2022} (bottom panel) are compared with the predictions of our chemical evolution models (see \citealt{ji2022} for a further 32 red giant stars). The observed and predicted histograms in Fig. \ref{fig:normal_cmd_comp} are normalized to unity. Even though the observed data are based on a limited spectroscopic sample of red giant targets, our models can reproduce the range of [Fe/H] abundances as covered by observations. The MDF of both model galaxies is from the full sample of stars. Both models have an MDF peak $\approx$ 0.4 dex which suggests the models are slightly metal-poor with respect to the observed [Fe/H] abundance ratio. One possible cause for this slight discrepancy is the initial burst of gas accretion not high enough for the stars to become enriched quicker. Alternatively, $\tau_{\text{min,Ia}}$ could be reduced resulting in an earlier enrichment of iron and shift the MDF peak.

The second peak predicted in both models is the consequence of the minimum time delay ($\tau_{\text{min,Ia}}=152\,\text{Gyr}$) and normalisation constant ($N_{\text{Ia}}=8.0\times 10^{-3}\,\text{M}_{\sun}^{-1}$) of SNe Ia, resulting in a bimodal MDF. Perhaps a lower minimum delay time for SNe Ia is needed to mimic the observed MDF of stars in Car II and Ret II. A relatively higher normalisation constant in dwarf galaxies seems to be suggested by recent observational surveys (see \citealt{brown2019,Johnson2022}).

\begin{figure*}
    \centering
    \includegraphics[width = 1.00\textwidth]{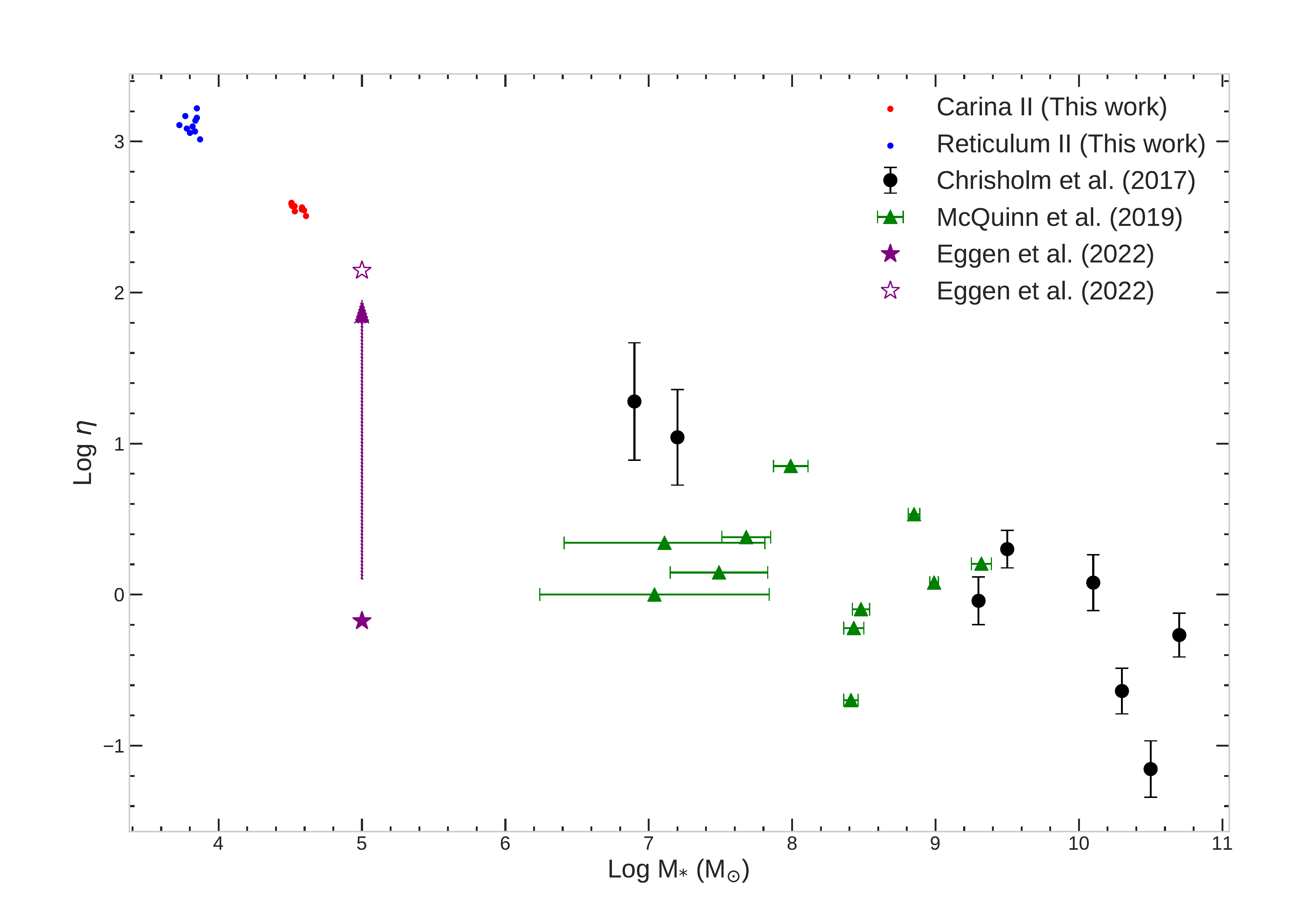}
    \caption{Average mass loading factor, $\eta$, as predicted in Car II (red points) and Ret II (blue points) as a function of total stellar mass at the present time. We run both models ten times, calculating $\eta$ along with the final stellar mass at the end of each simulation. The black circles correspond observations of \citet{Chrisholm2017} from the Cosmic Origins Spectrograph (COS; \citealt{Green2012}); the green triangles are from \citet{Mcquinn2019}, and the present-day mass loading factor as measured in the dwarf starburst galaxy Pox 186 corresponds to the filled star in purple, with the empty star in purple showing the mass loading factor necessary to remove $\text{H}_{\text{I}}$ gas mass as expected from galaxy scaling relations (see \citealt{Eggen2022} for more details).}
    \label{fig:mlf}
\end{figure*}

Figure \ref{fig:TGM} shows the predicted evolution of the stellar mass (blue solid line) and gas mass (red solid line) as a function of time for Car II (bottom panel) and Ret II (top panel). The gas content in Car II has a faster and more intense depletion than in Ret II due to a larger number of SN events occurring, which drive the galactic outflows (see Table \ref{table:prop}). The fluctuations in the total gas mass are due to the SN explosions within the model, which return gas mass back to the ISM, causing short-lived variations of the gas content as a function of time. The total gas mass left at the end of the simulation is $\approx 10^{3}\,\text{M}_{\odot}$ for Car II and Ret II. Additional mechanisms of further gas removal can be attributed to ram pressure stripping and reionization, which are not included in our chemical evolution model. Interestingly enough, the temporal evolution of the stellar mass in both galaxies shows similar features, with the declining trend in stellar mass at late times being determined by stars of all masses evolving off the main sequence.

Our models accurately predict the stellar mass of Car II, but overestimate the mass of Ret II by almost a factor of $10$, even though the observed estimates of \citet{Sales2017} are affected by large systematic uncertainty. The total stellar number of stars is indicative of the uncertainty surrounding them within UFDs. As such, these numbers (see Table \ref{table:prop}) are taken as a rough estimate as our galaxy is modelled as a closed system with no external factors affecting its stellar structure.

\citet{McQuinn2023} examined the formation history and quenching of the UFD galaxy Pegasus W, located between the MW and M31 systems. They find that Pegasus W follows a similar quenching timescale to six other UFDs of the MW where galaxies formed 100\% of their stars by $\sim$ 11.6 Gyr. Even though reionisation is often attributed to the quenching of MW UFDs, environmental effects may have also played a part. Our models not only attribute quenching to SN ejecta but also the reason why we do not observe gas in these systems.

\subsection{Estimating the mass loading factor}

One of the key aspects of galaxy formation and evolution studies is the relationship between the outflow rate and the SFR, and how this relationship changes as a function of the stellar mass to explain the average metallicity of galaxies. The mass loading factor, $\eta$, is defined as the ratio between the outflow rate and the SFR by means of the following equation:

\begin{equation}
\label{massloading}
   \Big(\frac{\text{d}\,\rho_{\text{gas}}}{\text{dt}}\Big)_{\text{outflow}} \, = \eta \cdot \text{SFR(t)},  
\end{equation}

\noindent where the left-hand side, $\big(\text{d}\rho_{\text{gas}}/\text{dt}\big)_{\text{outflow}}$, corresponds to the gas ejected out of the galaxy through SN feedback per unit time, and $\text{SFR}(t)$ is the star formation rate at the time $t$. 
We can use equation \ref{massloading} to compute the mass loading factor, $\eta$, as the simulated galaxy evolves as a function of time. We note that equation \ref{massloading} can only be used when there is star formation activity within the galaxy. Galactic winds can, however, still develop in our simulation even when there is no star formation activity within the galaxy, as they can be efficiently driven by SN Ia explosions.  

Figure \ref{fig:mlf} shows the average mass loading factor, $\eta$, as predicted by our models for Car II (red points) and Ret II (blue points) as a function of their stellar mass at $13.8\,\text{Gyr}$ (present-day). We run both models $10$ times -- by assuming the exact same parameters -- and, for each run, we take the average mass loading factor over the period of star formation. Ret II has higher $\eta$ values than Car II due to a lower SFR and less dense ISM, which facilitates the expansion of SN bubbles. In Figure \ref{fig:mlf}, we also show some measurements of the mass loading factor over a wide range of stellar masses from \citet[black points]{Chrisholm2017}, \citet[green triangles]{Mcquinn2019} and \citet[purple star symbol]{Eggen2022}. High mass galaxies tend to have a lower mass loading factor indicating that either they have a high SFR and/or a low outflow rate. The observed data from \citet[filled star symbol in purple]{Eggen2022} corresponds to the dwarf starburst galaxy Pox 186, which is observed to have relatively suppressed outflow at the present time as previous SN-driven outflow activity had efficiently removed large amounts of gas from the galaxy; the empty star symbol in purple corresponds to an estimate of the mass loading factor associated with the previous outflow episode in Pox 186, by estimating the total amount of lost gas (see \citealt{Eggen2022} for more details). 

Our Car II and Ret II models have an average mass loading factor of $364$ and $1304$, respectively, which are computed by averaging out 10 simulation runs. We note that our Ret II model is characterised by a smaller SFR than Car II, which results in a larger dispersion in both stellar mass and average mass loading factor. The values of $\eta$ that we predict in Car II and Ret II qualitatively align with the observed trend at high stellar masses, showing that galactic winds are increasingly important in the chemical evolution of galaxies as the stellar mass of the system becomes increasingly smaller, being crucial to shaping the observed mass-metallicity relation of galaxies (e.g., \citealt{spitoni2010,spitoni2017}). 

\section{Chemical abundance patterns}
\label{sec:ce}

In this Section, we show and discuss the predictions of our reference inhomogeneous chemical evolution models for the chemical abundance patterns in Car II and Ret II. 

\begin{figure*}
    \centering
    \includegraphics[width = 0.99\textwidth]{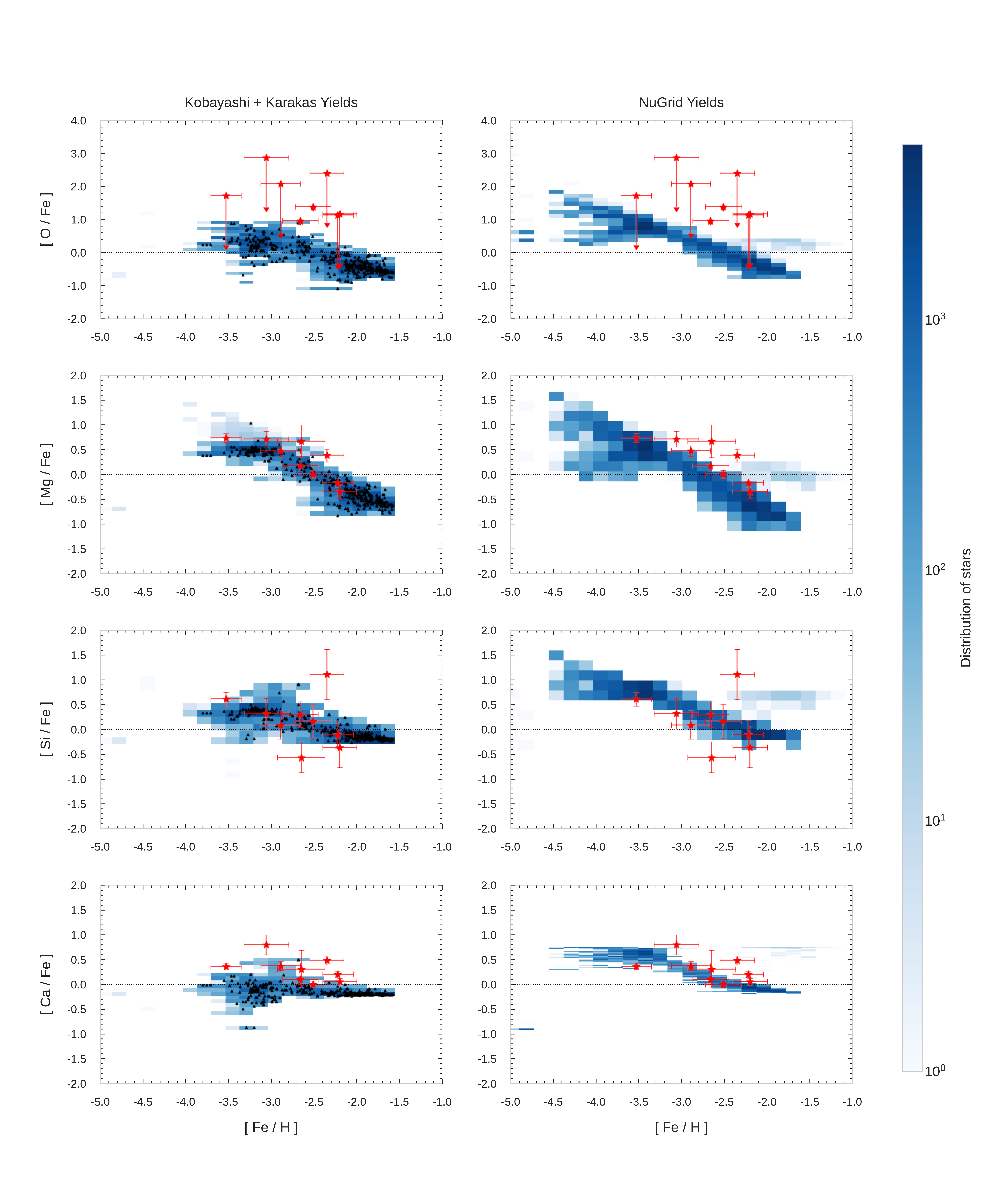}
    \caption{\textit{Left panels} -- [X/Fe]-[Fe/H] abundance diagrams as predicted by our reference chemical evolution model for Car II assuming the stellar yields \citet{Kobayashi2006} for massive stars and those of \citet{Karakas2010} and AGB stars. From top to bottom, the various panels show our results for O, Mg, Si, and Ca. Model predictions are shown as 2D histograms with a colour map corresponding to the number of stars within each grid element, while the filled stars in red are the observations from \citet{Ji2020}. \textit{Right panels} -- The same as the left panels, but assuming the NuGrid stellar yields \citep{Ritter2018}. The black triangles are the RGB stars within our model as determined in Section \ref{sec:cmd}.}
    \label{fig:CarII_CA_alpha}
\end{figure*}

\subsection{Carina II}

\begin{figure*}
    \centering
    \includegraphics[width = 1.00\textwidth]{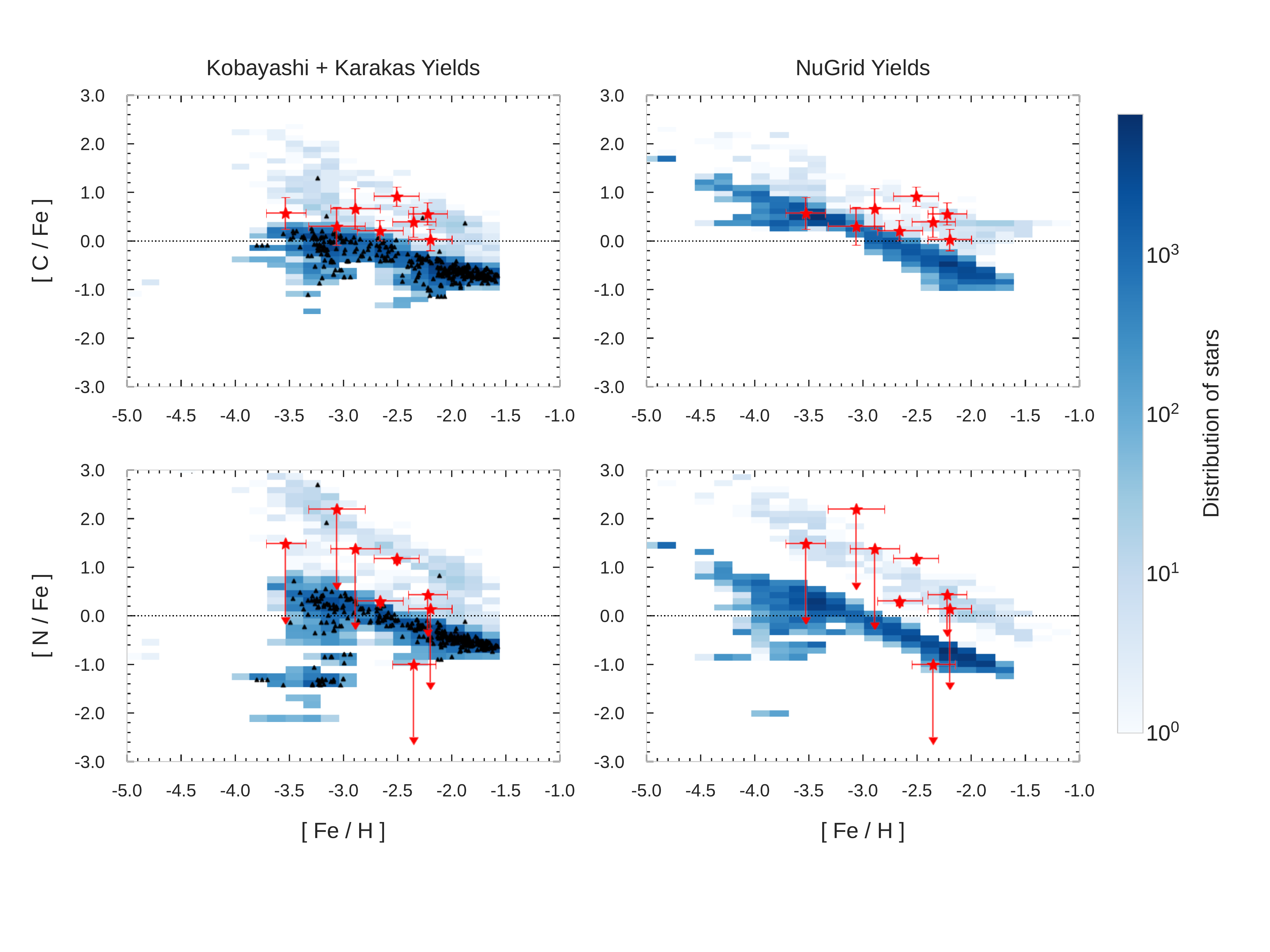}
    \caption{Same as Figure \ref{fig:CarII_CA_alpha}, but for C and N.}
    \label{fig:Car_II_CN}
\end{figure*}

\begin{figure*}
    \centering
    \includegraphics[width = 1.00\textwidth]{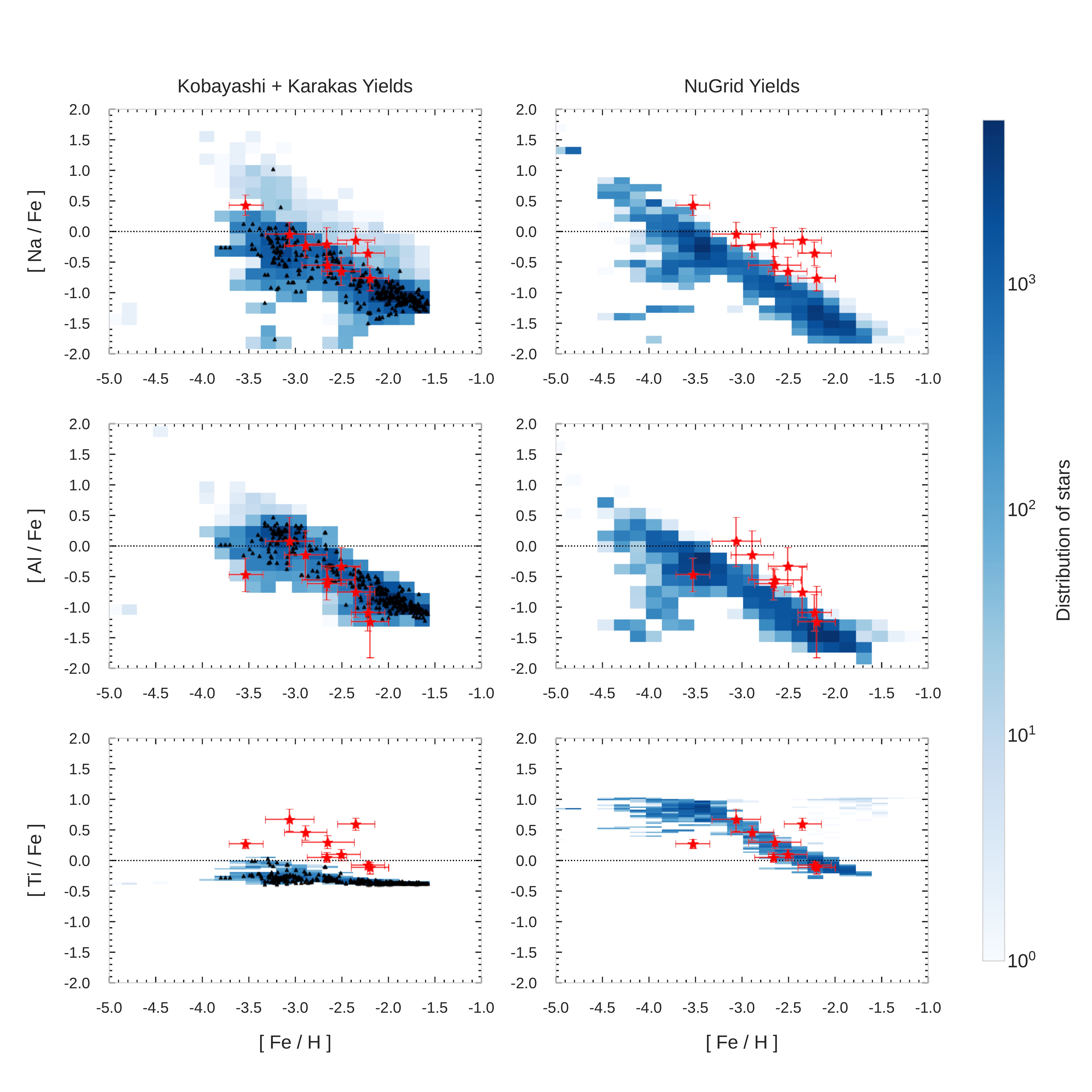}
    \caption{Same as Figure \ref{fig:CarII_CA_alpha}, but for Na, Al and Ti}
    \label{CarII_CA_odd}
\end{figure*}

Figure \ref{fig:CarII_CA_alpha} shows a comparison between the predicted [$\alpha$/Fe]-[Fe/H] abundance diagrams in Car II and the observations from \citet[filled red stars]{Ji2020}, where $\alpha$ corresponds to the $\alpha$-elements O, Si, Mg, and Ca. The model predictions in Fig. \ref{fig:CarII_CA_alpha} are shown as 2D-histograms with the colour corresponding to the number of stars within each grid element. The black triangles in each diagram show the predicted abundances in the model RGB stars, as obtained by coupling our model with a set of stellar evolutionary tracks (see Section \ref{sec:cmd} for more details). The panels on the left show the predictions of our model assuming the stellar yields of \citet{Kobayashi2006} for massive stars and \citet{Karakas2010} for AGB stars, whereas the panels on the right show our predictions when assuming the NuGrid stellar yields \citep{Ritter2018}. 

In Fig. \ref{fig:CarII_CA_alpha}, O and Mg represent the $\alpha$-elements that are mostly synthesized by CCSNe, whereas the nucleosynthesis of Si and Ca also have a non-negligible contribution from SNe Ia. The predicted trend of [$\alpha$/Fe]-[Fe/H] in Fig. \ref{fig:CarII_CA_alpha} stems from the so-called \enquote*{time-delay model} \citep{Tinsley1980,matteuccigreggio1986,matteuccibrocato1990}, being due to the low star formation efficiencies assumed in Car II, which cause Type Ia SNe to explode and pollute the ISM with iron at very low [Fe/H] abundances ($[\text{Fe/H}]\approx -3.0$ in the model). Before the first SNe Ia explode (namely, when $[\text{Fe/H}] \lesssim -3.0$), [$\alpha$/Fe] only stems from the chemical enrichment of massive stars; then, after the first SNe Ia explode, [$\alpha$/Fe] decreases because of the large amount of Fe produced and ejected by SNe Ia, causing a \enquote*{knee} in [$\alpha$/Fe]-[Fe/H]. In our Car II model, SNe Ia are predicted to occur after $\tau_{\text{min,Ia}}=152 \,  \text{Myr}$, releasing iron into the galaxy ISM and causing the [$\alpha$/Fe] of the subsequent stellar generations to be increasingly lower as a function of time.  A more intense star formation activity would determine a larger amount of iron from core-collapse SNe, and -- as a result of this -- the decrease of [$\alpha$/Fe] as due to SNe Ia would happen at higher [Fe/H]. For this reason, [$\alpha$/Fe] can be thought of as a sort of \enquote*{chemical clock} that measures the level of star formation activity within galaxies. 

The majority of [O/Fe] abundance measurements in Car II from \citet{Ji2020} are upper limits. Our models always lie well below the observations of \citet{Ji2020}, predicting [O/Fe] ratios at $[\text{Fe/H}] \lesssim -3.0$ that are in line with the abundance measurements in Galactic halo stars (e.g., \citealt{Gratton2003,Akerman2004,Cayrel2004}; see also Figure 1 in \citealt{vincenzo2017}). Similarly to [O/Fe], the main contributors of Mg are CCSNe, which cause [Mg/Fe] to be constant at $[\text{Fe/H}] \lesssim -3.0$. We can also use [Mg/Fe] to predict the enrichment timescale by observing stars at various [Fe/H] and finding a \enquote*{knee}. However, from the observations of \citet{Ji2020}, it is unclear where the \enquote*{knee} took place in the [Mg/Fe]-[Fe/H] diagram. There is a difference in [Mg/Fe]-[Fe/H] between our model with the \citet{Kobayashi2006} yields and that with the Nugrid yields, with the \enquote*{knee} being clearly visible only when assuming the \citet{Kobayashi2006} yields. A similar difference between the two models is also seen in the predicted [O/Fe]-[Fe/H] abundance diagram.

At variance with [O/Fe] and [Mg/Fe], the [Si/Fe]-[Fe/H] abundance pattern as predicted by our model with the \citet{Kobayashi2006} massive star yields does not show a clear \enquote*{knee} at the onset of SN Ia enrichment, which is seen instead by the model with the Nugrid yields. It is worth noting that the spread in [Si/Fe] as predicted with the stellar yields of \citet{Kobayashi2006} is higher than that predicted when assuming the Nugrid yields. Despite being based on different stellar models, both the model assuming Nugrid yields and that with \citet{Kobayashi2006} yields can reproduce the majority of [Si/Fe] measurements in Car II.  The final diagram in Fig. \ref{fig:CarII_CA_alpha} is [Ca/Fe]-[Fe/H], in which both models underestimate the bulk of observational data. Interestingly, similarly to what we find for [Si/Fe]-[Fe/H] (both Si and Ca have a non-negligible contribution from SNe Ia), the model assuming the Nugrid yields predicts a clear \enquote*{knee} in [Ca/Fe]-[Fe/H], which is not seen in the model with the \citet{Kobayashi2006} yields. 

\begin{figure*}
    \centering
    \includegraphics[width = 0.99\textwidth]{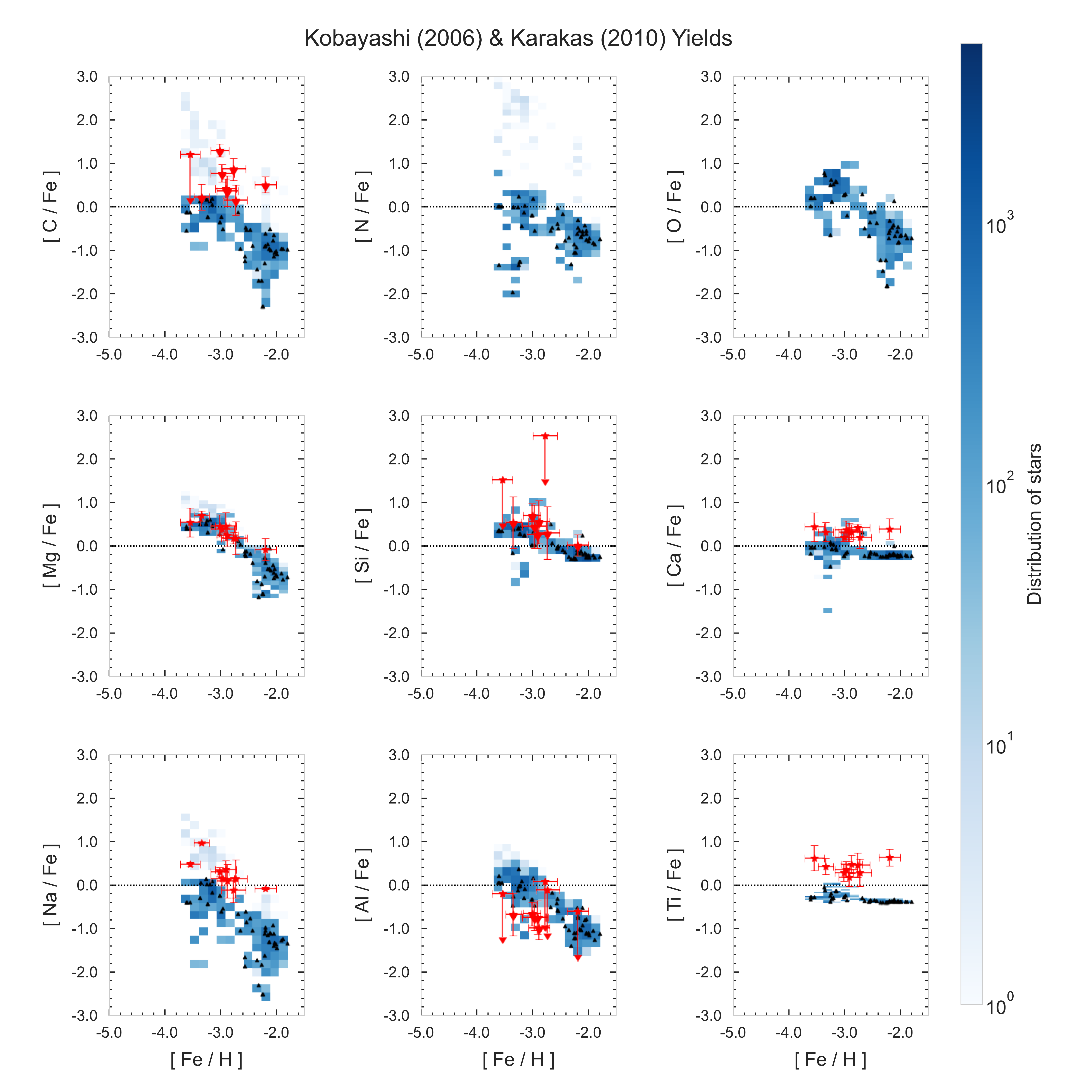}
    \caption{[X/Fe]-[Fe/H] abundance diagrams as predicted by our reference chemical evolution model for Ret II assuming the stellar yields \citet{Kobayashi2006} for massive stars and those of \citet{Karakas2010} for AGB stars. Observations in Ret II from \citet{Ji2020} are shown as filled stars in red. Similar to Fig. \ref{fig:CarII_CA_alpha}, \ref{fig:Car_II_CN}, and \ref{CarII_CA_odd} are the black triangles depicting the red giants stars within our Ret II model.}
    \label{fig:RetII_KK}
\end{figure*}

\begin{figure*}
    \centering
    \includegraphics[width = 0.99\textwidth]{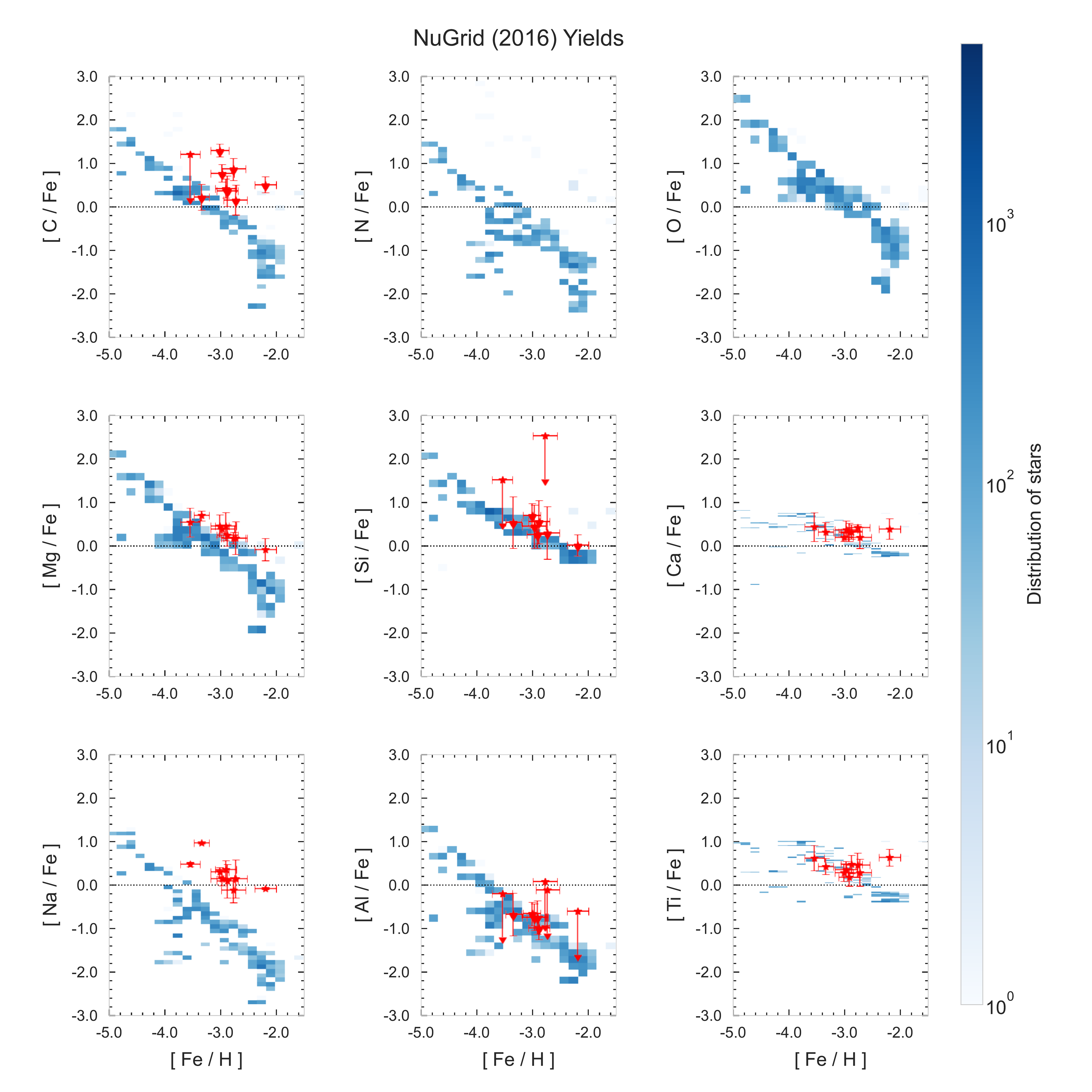}
    \caption{Similar to Figure \ref{fig:RetII_KK}, but for NuGrid \citep{Ritter2018} yields.}
    \label{fig:RetII_NG}
\end{figure*}

Figure \ref{fig:Car_II_CN} shows the predictions of our chemical evolution models for [C/Fe] and [N/Fe] as a function of [Fe/H]. The  predicted [C/Fe]-[Fe/H] abundance diagram is characterized by a sequence of C-rich stars -- with [C/Fe] above solar -- at all metallicities. They originate from isolated chemical enrichment events associated with CCSNe and low-mass AGB stars (with mass in the range $1\lesssim M/\text{M}_{\sun} \lesssim 4$), which provide an enhanced enrichment for stars forming in their debris. Interestingly enough, several observed stars in Car II can be classified as Carbon-Enhanced Metal-Poor (CEMP) stars with $\text{[C/Fe]} \ge 0.7$ at $[\text{Fe/H}] \leq -1$ \citep{Jeon2021,Myoungwon2021}. Overall, our models underestimate [C/Fe] in Car II. We also note that C is systematically depleted in the atmosphere of RGB stars as a consequence of the first dredge-up, with the observed C abundances as measured at the present time in the atmosphere of RGB stars being lower than the C abundances at birth (e.g., see figure 16 of \citealt{Norris2010} for a study of how the stellar abundances of C vary as a function of stellar luminosity in a sample of stars within UFDs that have larger total stellar mass than Car II; see also the analysis of \citealt{vincenzo2021} and references therein). Accounting for the variation of C abundances due to the first dredge-up would make the comparison with the model predictions worse. There are studies in the literature suggesting that CEMP stars in UFD galaxies result from the chemical enrichment of Pop III stars (e.g., \citealt{Salvadori2012,Ji2020,Jeon2021}). Even though our models with the Nugrid yields produce some CEMP stars at extremely low metallicity, they do not reproduce the [C/Fe] observations in Car II. Conversely, the models assuming the \citet{Kobayashi2006} yields do not predict any CEMP stars at extremely low metallicity. In Figure \ref{fig:Car_II_CN}, we also show [N/Fe]-[Fe/H], in which most observational data are upper limits. The predictions of both models are consistent with the measured [N/Fe] upper limits. However, we caution the readers that N abundances are also affected by the first dredge-up in RGB stars, being enhanced in the stellar atmosphere after RGB stars go through the first dredge-up (e.g., see \citealt{vincenzo2021} and references therein).  

Figure \ref{CarII_CA_odd} shows the predictions of our chemical evolution models for two odd-Z chemical elements, Na and Al, along with Ti. The observed [Na/Fe]-[Fe/H] and [Al/Fe]-[Fe/H] abundance patterns are reproduced by our model with the \citet{Kobayashi2006} yields for massive stars and the \citet{Karakas2010} yields for AGB stars. Interestingly, the predicted [Na/Fe]-[Fe/H] diagram shows a similar feature as the [C/Fe]-[Fe/H] and [N/Fe]-[Fe/H] diagrams (see Figure \ref{fig:Car_II_CN}), predicting a sequence of [Na/Fe]-rich stars, which stems from the chemical enrichment of isolated high mass CCSNe progenitors and low-mass AGB stars. The model with the \citet{Kobayashi2006} stellar yields severely underestimates the [Ti/Fe] abundance ratios, which are instead better reproduced by the model with the Nugrid stellar yields. \citet{Kobayashi2006} suggested that a \enquote*{jetlike} explosion with high entropy could be required to produce enhanced Ti yields in their stellar models. In the Nugrid models \citep{Ritter2018}, the main massive-star producers of Ti have masses in the range $15 < M / \text{M}_{\sun} <  20$.

To summarise, the predicted chemical evolution of Car II is very dependent on the assumed set of stellar nucleosynthetic yields. When assuming the stellar yields of \citet{Kobayashi2006} for massive stars and the stellar yields of \citet{Karakas2010} for AGB stars, we can qualitatively reproduce the observed [Mg/Fe] and [Si/Fe] ($\alpha$-elements) as well as [Na/Fe] and [Al/Fe] (odd-Z elements). The situation with [Ti/Fe] is very different, as the model assuming the Nugrid yields produces better results when compared to observational data. [C/Fe] ratios are severely underproduced, suggesting that a different yield set is needed. We also caution the readers that C abundances in RGB stars are affected by the first dredge-up, which causes a depletion of the surface abundances of C with respect to those at birth, making the comparison to the model predictions worse. 

\subsection{Reticulum II}

Figure \ref{fig:RetII_KK} shows the results of our reference model for Ret II, in which the observed [$X$/Fe] vs [Fe/H] diagrams from \citet[filled red star symbols]{Ji2016} are compared with the predictions of our models, with $X$ corresponding to C, Mg, Si, Na, Al, and Ca. We only show the predictions of the chemical evolution model assuming the stellar yields of \citet{Kobayashi2006} for massive stars and \citet{Karakas2010} for AGB stars. The predictions of the model are shown as a 2D histogram with the colour coding corresponding to the number of stars in each bin, where the majority of the stars are found in darker areas of the panels. The black triangles correspond to the predicted chemical abundances in the model RGB stars. 

The best match to observational data is obtained for the [Mg/Fe]-[Fe/H] and [Al/Fe]-[Fe/H] abundance diagrams, in which we predict a \enquote*{knee} at $[\text{Fe/H}] \approx -3$, corresponding to the time when the first SNe Ia are expected to explode in the system. Our model for Ret II assumes a minimum delay time for SNe Ia $\tau_{\text{min,Ia}}=152\,\text{Myr}$ (see Table \ref{table:prop}). Also, the majority of [Si/Fe] ratios are reproduced by the model with the exception of one star at $[\text{Fe/H}]\approx -3.6$, which shows a strong enhancement in [Si/Fe] that cannot be explained by the model. The [Na/Fe], [Ca/Fe], and [C/Fe] abundance ratios are underestimated by our chemical evolution model, similar to what we found for Car II.

Interestingly enough, the model predicts the presence of red giant stars across the given metallicity range, where approximately 60\% of these stars were iron enhanced from SNe Ia events. Several stars appear to have comparable chemical abundance patterns to observations suggesting perhaps that their formation history could be similar. 

Recent works of \citet{Ji2020} and \citet{ji2022} proposed that the chemical evolution of Ret II might have resulted from two or more periods of star formation, which was suggested to explain the observed r-process elemental abundances. Our reference model for Ret II only has a single infall of gas which produces a single continuous episode of star formation, recreating the observed [Mg/Fe], [Si/Fe], and [Al/Fe] of the stars along with the predicted abundance spread. In the next Section, we also show that our model for Ret II can qualitatively recreate the observed CMD.

Similar to Fig. \ref{fig:RetII_KK}, Figure \ref{fig:RetII_NG} shows the chemical abundances of 9 elements for Ret II with the stellar yields of NuGrid \citep{Ritter2018}. We exclude RGB stars in this Figure as their abundance pattern is eerily similar to those depicted in Fig. \ref{fig:RetII_KK}. The abundance ratios [Si/Fe], [Ca/Fe], [Al/Fe] and [Ti/Fe] reproduce the observed chemical abundances, whereas the others are slightly overproduced with respect to [Fe/H].

\section{Colour Magnitude Diagrams}
\label{sec:cmd}

\begin{figure*}
    \centering
    \includegraphics[width = 1.00\textwidth]{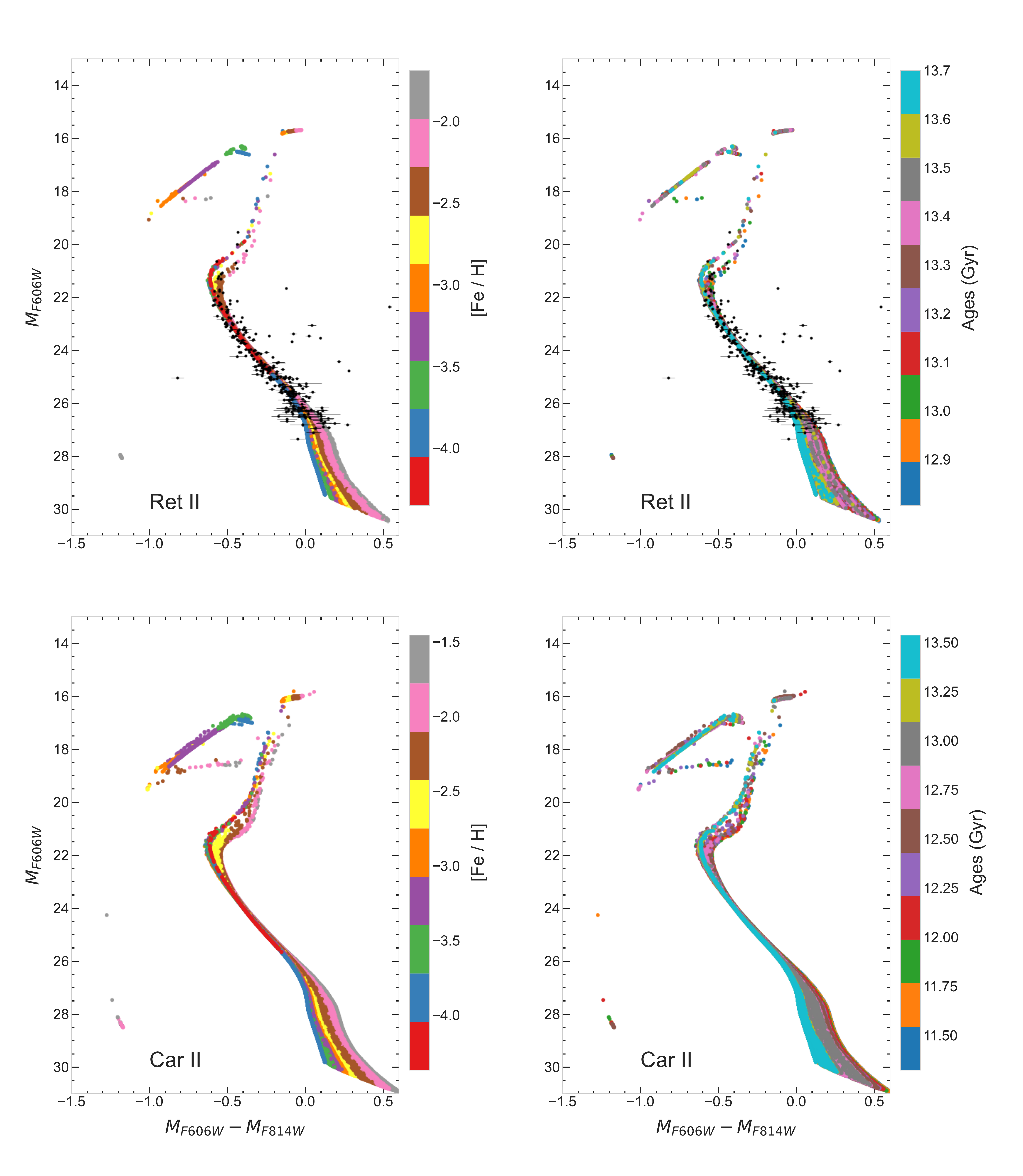}
    \caption{The figure shows the CMD as predicted by our models for Car II (lower panels) and Ret II (upper panels), by lighting up the model stars with different mass, age, and metallicity, adopting the MIST isochrone database \citep{Dotter2016,Choi2016}. In the panels on the left, the colour scheme corresponds to the iron abundance, $[\text{Fe/H}]$, of the model stars; in the panels on the right, the colour scheme corresponds to the ages of the model stars. The black points with error bars in the top panel are observed stars in Ret II from \citet{sacchi2021}.}
    \label{fig:RetII_CMD}
\end{figure*}

One of the main appeals of {\tt i-GEtool} is that we can create a population of synthetic stars, each with a mass, metallicity and age, which result from the Monte Carlo sampling of the IMF as the galaxy forms stars and undergoes chemical enrichment, by accounting for the stellar lifetimes as stars evolve off the main sequence; this allows us to create a synthetic CMD, starting from a database of stellar evolutionary tracks. CMDs are routinely used to reconstruct the SFH and age-metallicity relations in UFDs, by fitting the position of the stars in the observed CMD with stellar evolutionary tracks (e.g., see \citealt{Dolphin2002,Cignoni2010,Monelli2010,deboer2011,Weisz2014,Gallart2021}); here we take the opposite approach, as we start from the predictions of our inhomogeneous chemical evolution model and use the observed CMD as an observational constraint along with the observed chemical abundance patterns (see also \citealt{Vincenzo2016} for an example of this approach for the Sculptor dwarf spheroidal galaxy). In this section, we explain how we acquire luminosity, temperature, and photometric properties from our model stars. Afterwards,  we show the results of our analysis, comparing the synthetic CMD with the observed ones in Car II and Ret II.

\subsection{Interpolating Isochrones}

In our method, we use the metallicity, mass and age of each star formed by our inhomogeneous chemical evolution model to retrieve its synthetic photometric properties within the MIST isochrone database\footnote{The MIST isochrone database for different synthetic photometric systems can be downloaded at the following link: \url{https://waps.cfa.harvard.edu/MIST/}.} \citep{Dotter2016,Choi2016}, which are based on the MESA stellar evolution code \citep{MESA2011,MESA2013,MESA2015,MESA2018}. We adopt isochrones computed with stellar models with rotation velocity $v_{\text{rot}}/v_{\text{crit}} = 0.4$, where $v_{\text{crit}}$ corresponds to the critical break-up velocity, as stars are observed to rotate faster at the low metallicities of UFD galaxies (e.g., see \citealt{Amard2020}, who analysed the distribution of rotation velocities as a function of mass and metallicity in a sample of MW stars in the \textit{Kepler} fields). We use all available isochrones in the MIST database, that span iron abundances in the range $-4 \leq [\text{Fe/H}] \leq 0.5$ and ages in the range $5 \leq \log\big( \text{age}/\text{yr} \big) \leq 10.3 $. Linear interpolation is used to obtain the luminosity, temperature, and other photometric properties of the synthetic stars along the isochrone tracks. 

In our analysis of Ret II, since we compare our models to the HST observations of \citet{sacchi2021} \footnote{Even though our analysis of Ret II was conducted with the CMD from \cite{sacchi2021}, a larger catalogue can be found in \cite{Simon2022} where they covered $<$ 10 \% of Ret II.}, the results for the synthetic CMD are shown by focusing on the HST bands F606W and F814W; for Car II, we show two synthetic CMDs, the first using the HST bands F606W and F814W and the second using the DECAM $g$ and $r$  bands to make a comparison with the observations of \citet{Li2018}. 

Our results for the CMDs of Ret II and Car II using the HST bands F606W and F814W are shown in Figure \ref{fig:RetII_CMD}. In the top panels, we compare the synthetic and observed CMD of Ret II. In the bottom panels, we show our predictions for Car II. The black points with error bars in the top panels correspond to the HST-observed stars of \citet{sacchi2021} in Ret II. The colour scheme in the left panels corresponds to the iron abundances, $[\text{Fe/H}]$, of the synthetic stars, whereas the colour scheme in the right panels shows the ages of the synthetic stars.

\begin{figure}
    \centering
    \includegraphics[width = 0.48\textwidth]{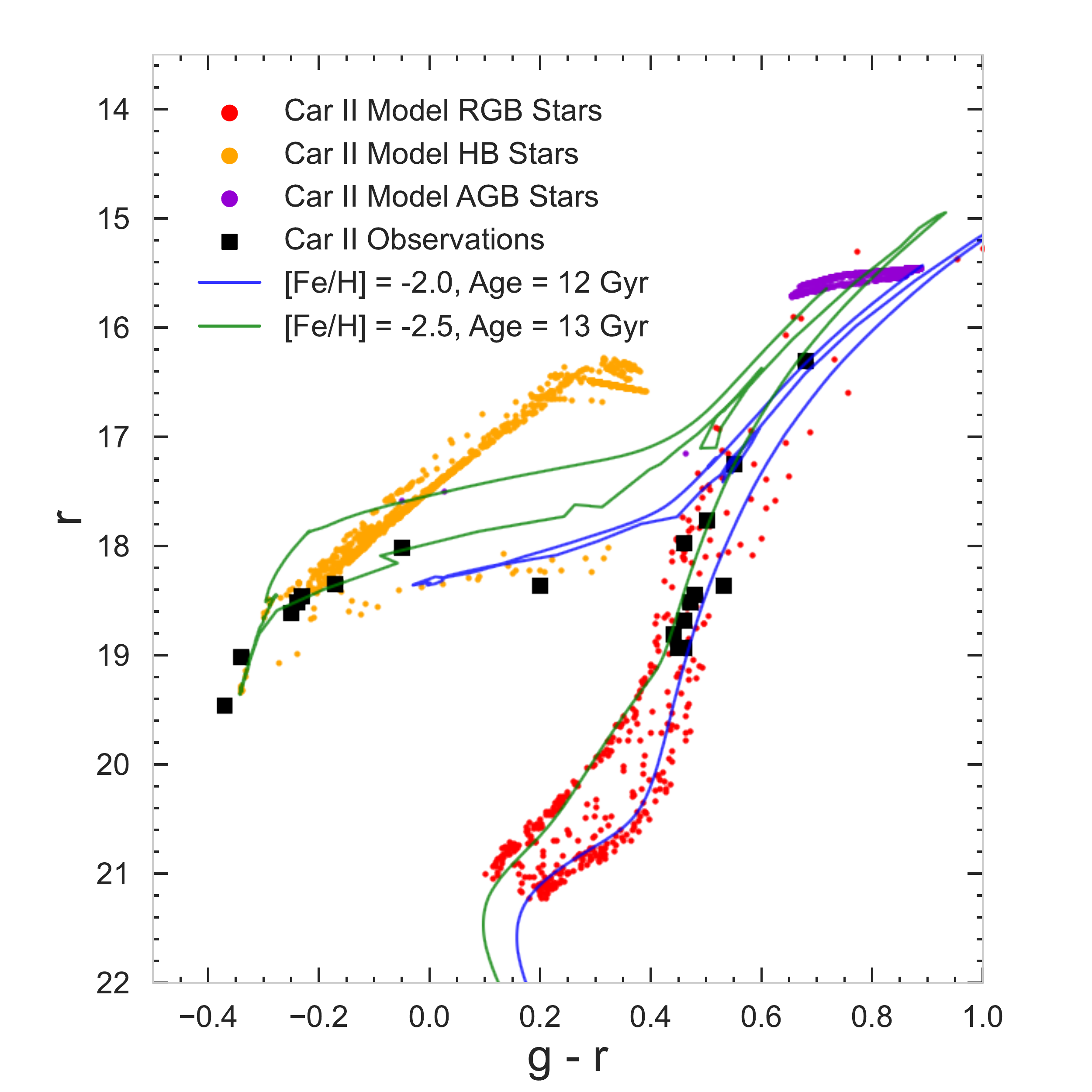}
    \caption{A direct comparison between the Car II observations from \citet[black squares]{Li2018} and the model post main sequence stars (red, yellow and purple points), resulting from the Monte Carlo sampling of the IMF while the galaxy undergoes star formation and chemical enrichment, by assuming the MIST isochrone database \citep{Dotter2016,Choi2016}. The blue solid line shows the predictions of a MIST isochrone with iron abundance $\text{[Fe/H]}=-2.0$ and age $13\,\text{Gyr}$, which characterises most red giants in our Car II model; the green solid line illustrates the effect of changing the iron abundance by $0.5\,\text{dex}$ on the predictions of stellar evolution models, showing a MIST isochrone with iron abundance $\text{[Fe/H]}=-2.5$ and age $13\,\text{Gyr}$.}
    \label{fig:CarII_CMD}
\end{figure}

\subsection{Ret II CMD}

In Fig. \ref{fig:RetII_CMD}, we assume that Ret II has a distance modulus $17.50$ from \citet[Table 1]{sacchi2021}, accounting also for the different zero-point values between the ACS HRC photometric system of the model isochrones and the ACS WFC system of the observations. Finally, we correct for the reddening and dust extinction according to \citet[Table 1]{sacchi2021} and \citet[Table 6]{Schlegel1998} using the \citet{Green2018} dust maps, assuming $R_{\text{V}}=3.1$. Similar corrections are applied for Car II in the bottom panel of Fig. \ref{fig:RetII_CMD}, assuming a distance modulus $17.87$ \citep{Li2018}.

When comparing the predictions of our model to observational data, the model slightly overpredicts the turn-off, which resides at $M_{\text{F606W}} - M_{\text{F814W}}= -0.5$, with the vast majority of stars being on the MS. Our Ret II model predicts a distribution of RGB, horizontal branch (HB), asymptotic giant-branch (AGB) and white dwarf (WD) stars at various metallicities and ages. There are two isochrones which dominate the CMD, the first having a population of metal-rich and young stars and the other dominant in metal-poor and old stars, which is captured by our inhomogeneous chemical evolution model, as suggested by \citet{Ji2020} and \citet{ji2022}. We note, however, that the synthetic CMD is not convolved with the observed photometric errors, which become increasingly larger at fainter luminosities, hence the synthetic MS has a lower spread than the observed one. The synthetic CMD for Car II in the bottom panel of Fig. \ref{fig:RetII_CMD} predicts a more populated RGB and horizontal branch, which is due to a more vigorous SFH than in Ret II.

As our model is based on the Monte Carlo random sampling of the \citet{Kroupa2001} IMF, the number of stars in each evolutionary phase varies from model to model. The following are the number of stars in each phase of the Ret II model in our present work:

\begin{enumerate}

    \item Main Sequence Stars: 23,360 
    \item Red Giant Branch Stars: 64
    \item Horizontal Branch Stars: 452
    \item Asymptotic Giant Branch Stars: 206
    \item White Dwarf Stars: 39

\end{enumerate}

The vast majority of our stars are on the main sequence with $m \leq$ 0.8 M$_{\odot}$ at various ages and metallicities. Very limited number of RGB stars within the stellar mass limit 0.75 M$_{\odot} \leq m \leq$ 0.82 M$_{\odot}$ are present in our favoured Ret II model. Interestingly the HB stars are the second most dominant evolutionary phase in our model over the larger mass range 0.78 M$_{\odot} \leq m \leq$ 0.93 M$_{\odot}$. Even with a lower number of AGB and WD stars in the model, they are still present at similar mass ranges. WD stars are considerably more metal rich at [Fe/H] $\geq$ -2, AGB stars have a metallicity range -3 $\leq$ [Fe/H] $\leq$ -2, while RGB and HB stars have a metallicity range -4 $\leq$ [Fe/H] $\leq$ -2.

\subsection{Car II CMD}

In Figure \ref{fig:CarII_CMD}, we compare Car II observations from \citet[black squares, with DECAM photometry]{Li2018} with our synthetic CMD (red, yellow and purple points), in which we only show stars above the main-sequence turn-off. Red giant stars in our Car II model can be characterized by the two isochrones, the first with age $12 \, \text{Gyr}$ and iron abundance $\text{[Fe/H]} = -2$ (blue solid line in the Fig. \ref{fig:CarII_CMD}) and second with age $13 \, \text{Gyr}$ and iron abundance $\text{[Fe/H]} = -2.5$ (green solid line). When comparing the horizontal branch to the given isochrones, we find that they follow the blue lines with a select few following the green. Interestingly, the AGB stars in our Car II model follow both isochrones lines with the majority in the green. We note that most red giants in the sample of \citet{Li2018} also have spectroscopic [Fe/H] abundance measurements, which cover values in the range $-3 \lesssim \text{[Fe/H]} \lesssim -2$. Observed stars from \citet{Li2018} follows similar isochrones to our Car II model, indicating a possible shared age and metallicity, and even perhaps formation history.

Similar to the above discussion, our present Car II model produces the following:

\begin{enumerate}

    \item Main Sequence Stars: 107,612 
    \item Red Giant Branch Stars: 382
    \item Horizontal Branch Stars: 1,018
    \item Asymptotic Giant Branch Stars: 1,142
    \item White Dwarf Stars: 671

\end{enumerate}

We found similar mass ranges and metallicities for our Car II RGB, HB, AGB and WD model stars with negligible variations. Even with different models, the metallicity range of these stars is identical to those in our Ret II model, due to the interpolation of the MIST isochrone database. Both models suggest that even with a single accretion event two isochrones can be derived including one with a young metal-rich population. 

Interestingly enough, we found a considerable number of HB stars within our models which are not observed in the night sky. The HST photometry and analysis conducted in \citet{sacchi2021} for Ret II did not contain any observed HB stars due to them being saturated in their images. \citet{Simon2022} found 2 HB stars in their DES photometry for Ret II where they added extra HST fields to get both of them in order to derive the distance. This is a possible indication that we are over-predicting the stellar mass of Ret II.

There is a superposition of stellar populations with similar ages and different metallicities resulting in two sequences within the HB phase. An analysis of this particular phase was conducted where we kept the metallicity and stellar age constant and varied the stellar mass for a star in order to test the sensitivity of the stellar mass on the isochrones. We found that the star traversed through the HB phase twice with the first having a very small mass range of order $m \sim 10^{-4}$ M$_{\odot}$. A slightly higher mass range would place the star more likely in the second phase in the HB sequence as it evolves back towards the RGB/AGB. In both Fig. \ref{fig:RetII_CMD} and \ref{fig:CarII_CMD}, the HB and AGB phases appear to have diagonal features due to the interpolation of the MIST isochrones, with each run of the model producing unique a chemical abundance and CMD.

\section{Conclusions}
\label{conclusion}

We have modelled two UFD galaxies, Car II and Ret II, which are satellites of the LMC and are on their first infall into the MW along with the LMC. In particular, we have developed  an inhomogeneous chemical evolution model, {\tt i-GEtool}, which assumes a single gas accretion event for the galaxy formation at early times, the IMF of \citet{Kroupa2001}, and a novel outflow method which is driven by SN explosions. The predictions of our models have been compared against the observed chemical abundances of \citet{Ji2016} for Ret II and \citet{Ji2020} for Car II, both taken with the MIKE spectrograph on the Magellan Clay telescope, by exploring two alternative sets of stellar nucleosynthetic yields: \textit{(i)} the first set assumes the stellar yields of \citet{Kobayashi2006} for massive stars and \citet{Karakas2010} for AGB stars, and \textit{(ii)} the second set assumes the stellar yields from the Nugrid collaboration \citep{Ritter2018}. 

From the masses, ages, and metallicities of the population of synthetic stars in our models, we could draw synthetic CMDs; for Ret II, we have compared the synthetic CMD against the HST observations of \citet{sacchi2021}, whereas for Car II we have used the observations of \citet{Li2018}, as obtained with the Magellan Baade Telescope, the Anglo-Australian Telescope, and the Very Large Telescope. Our results provide further insight into the SFHs and chemical evolution of Ret II and Car II to better understand how they may have formed and evolved. The following is a summary of the conclusions of our present work.

\begin{enumerate}

    \item Car II is expected by our model to have a more vigorous star formation history than Ret II, but both galaxies develop SN-driven galactic outflows at approximately the same time, which quenched the star formation by removing gas out of the galaxy. Galactic winds in Car II develop because of the relatively larger number of CCSNe and SNe Ia, whereas the lower gas densities in Ret II allow SN bubbles to expand more efficiently outside of the simulated galaxy volume, giving rise to galactic outflows of gas. The more intense star formation activity in Car II gives rise to higher total stellar masses as well as to a declining trend of [$\alpha$/Fe]-[Fe/H] which takes place at higher [Fe/H] than in Ret II. Finally, our model for Car II predicts a steeper stellar age-metallicity relation at early times than in Ret II, which is due to the more intense star formation activity in Car II, giving rise to a larger number of massive stars, exploding as CCSNe, at early times. 
    
    \item Our inhomogeneous chemical evolution models cover the range of [Si/Fe]-[Fe/H], [Mg/Fe]-[Fe/H], [Al/Fe]-[Fe/H], and [Na/Fe]-[Fe/H] as observed in the red giants of Ret II and Car II. We assumed a minimum delay-time for the onset of SNe Ia $\tau_{\text{min,Ia} }=152\,\text{Mr}$ for both Car II and Ret II. We note that the observed [Fe/H] distributions are based on a limited number of red giants (nine per galaxy), hence their peaks and widths might not be representative of the true stellar populations, also because of strong selection effects. 
    
    \item The stellar nucleosynthesis yields of \citet{Kobayashi2006} for massive stars overall provide the best qualitative agreement to observational data, predicting a larger spread of [$\alpha$/Fe] at fixed [Fe/H]. Most observations of [O/Fe] are upper limits, and the predictions of our models lie well below the observed values, with  [O/Fe] ratios from CCSNe enrichment agreeing with the measurements in Galactic halo stars (e.g., \citealt{Gratton2003,Akerman2004,Cayrel2004}). [Ca/Fe] and [Ti/Fe] are systematically underpredicted at all metallicities, whereas [C/Fe] and [N/Fe] are also underproduced. We discuss that the observed C and N abundances in the red giants of Car II and Ret II might be affected by internal mixing processes (first-dredge up; see \citealt{Norris2010,vincenzo2021}) which would reduce the surface C abundances in red giants (making the comparison to the model predictions worse) and increase the surface N abundances.
    
    \item Outflows from SN explosions can be used to explain the physical mechanism by which UFD galaxies had their SFHs quenched at early times, containing little traces of gas at the present time. Although there are other efficient methods of gas removal such as ram pressure stripping and reionization, which are not included in our model, we find that SN explosions alone -- by employing our method of modelling SN-driven galactic outflows -- allow the removal of the majority of the gas from UFD galaxies, in agreement with the findings of previous studies (e.g., \citealt{lanfranchi2003,lanfranchi2004,salvadori2009,vincenzo2014,romano2019,Gallart2021}). Some gas still remains within the galaxy but it can then be removed by the other aforementioned physical mechanisms.
    
    \item Ret II and Car II are characterized by high mass-loading factors, $\eta_{\text{RetII}} = 1304$ and $\eta_{\text{CarII}} = 364$, which qualitatively extend to lower stellar masses the trend of $\eta$ vs $M_{\star}$ which is seen at higher stellar masses. 
    
    \item When combining the predictions of our inhomogeneous chemical evolution model with the MIST database of isochrones \citep{Dotter2016,Choi2016}, the model for Ret II predicts similar numbers of red giants as in observations, but spanning the specific range of ages between $13.7\,\text{Gyr}$ and $12.8\,\text{Gyr}$ and metallicity -4 $\leq$ [Fe/H] $\leq$ -2, slightly over-predicting the colour index of the turn-off, which might also be a sign of a second, relatively metal-rich and younger stellar population in Ret II, as suggested by \citet{Ji2020,ji2022}. Also, our model for Car II predicts a similar number of red giants as in observations, but the position of the horizontal branch stars in the observed CMD can only be reproduced by including also a population of synthetic metal-poor horizontal-branch stars, with $[\text{Fe/H}]\approx -2.5$, which is produced by our model, as it predicts most synthetic horizontal-branch stars to have $[\text{Fe/H}]\approx -2$ when assuming the MIST isochrone database.

\end{enumerate}

In our future work, we plan to explore the chemical evolution and CMD of Eridanus II \citep{Gallart2021} and other recently discovered UFD galaxies of the MW, by including and investigating the effect of mixing in our model. Fluorine is understood to not have major contributions from novae \citep{Womack2023}, which could provide further insight into the evolution of this element in UFDs with our present model. Several components of nucelosynthesis can be understood in depth with {\tt i-GEtool} such as the stellar mass of supernovae progenitors, which can be used and adapted in future studies. Finally, we can generate artificial spectra from our interpolated isochrones to predict the spectral energy distribution of a model galaxy and compare it to observed spectra at high redshift.

\section*{Acknowledgements}

We thank David Weinberg for the valuable discussions. We acknowledge support from the Science and Technology Facilities Council (through the University of Hull’s Consolidated Grant ST/R000840/1) and the European Union’s Horizon 2020 Research and Innovation programme (ChETEC-INFRA -- Project no. 101008324). The authors further thank the {\tt viper} High-Performance Computing Facility team at the University of Hull for the allocation of computing time. The Ret II data are associated with the \textit{HST} Treasury Program 14734 (PI: Kallivayalil). Support for this program was provided by NASA through grants from the Space Telescope Science Institute.

%%%%%%%%%%%%%%%%%%%%%%%%%%%%%%%%%%%%%%%%%%%%%%%%%%
\section*{Data Availability}

The data underlying this article, which include the chemical abundances and the synthetic CMDs as predicted by our inhomogeneous chemical evolution models for Ret II and Car II, will be shared on reasonable request to the corresponding author.



\bibliographystyle{mnras}
\bibliography{References} % if your bibtex file is called example.bib

%\begin{thebibliography}{99}
%\end{thebibliography}

%%%%%%%%%%%%%%%%%%%%%%%%%%%%%%%%%%%%%%%%%%%%%%%%%%

%%%%%%%%%%%%%%%%% APPENDICES %%%%%%%%%%%%%%%%%%%%%

\appendix

%%%%%%%%%%%%%%%%%%%%%%%%%%%%%%%%%%%%%%%%%%%%%%%%%%

% Don't change these lines
\bsp	% typesetting comment
\label{lastpage}
\end{document}